%% file: neurips_2023.tex
\pdfoutput=1
\documentclass{article}
\input{PREFIX}

\title{CANDY: A Benchmark for Continuous Approximate Nearest Neighbor Search with Dynamic Data Ingestion}
 
\begin{document}
\input{authors}
\maketitle

\input{abstract}

\compact
\input{Introduction}

\compact
\input{BenchmarkSystem}

\compact

\input{ExperimentShort}
\compact
\input{Limitations}
\compact
\input{Conclusion}

\bibliographystyle{abbrv}
\bibliography{mybib}

\appendix
\input{Appendix}
\input{Evaluations/Extra}
\input{Evaluations/Setups}


\end{document}

%% file: PREFIX.tex
\usepackage{comment}

    \usepackage[preprint, nonatbib]{neurips_2023}

\usepackage[utf8]{inputenc} 
\usepackage[T1]{fontenc}    
\usepackage{hyperref}       
\usepackage{url}            
\usepackage{booktabs}       
\usepackage{amsfonts}       
\usepackage{nicefrac}       
\usepackage{microtype}      
\usepackage{xcolor}         
\usepackage{rotating} 
\usepackage{caption}

\hypersetup{
    colorlinks,
    linkcolor={red!50!black},
    citecolor={blue!50!black},
    urlcolor={blue!80!black}
}
\usepackage{graphicx}
\usepackage{xspace}
\usepackage{multirow}
\usepackage{subfigure}
\newcommand{\xianzhi}[1]{{\textcolor{purple}{#1}}}	
\newcommand{\notes}[1]{{\textcolor{red}{#1}}}	
\newcommand{\shuhao}[1]{{\textcolor{blue}{#1}}}	
\newcommand{\system}{\texttt{CANDY}\xspace}
\newcommand{\vectorSet}{\hyperref[tab:notations]{$S$}\xspace}
\newcommand{\vectorSubset}{\hyperref[tab:notations]{$C$}\xspace}
\newcommand{\resulSet}{\hyperref[tab:notations]{$O_{exact}$}\xspace}
\newcommand{\resulSetAKNN}{\hyperref[tab:notations]{$O_{aknn}$}\xspace}
\newcommand{\recall}{\hyperref[tab:notations]{$r$}\xspace}
\newcommand{\resulSetInEq}{\hyperref[tab:notations]{O_{exact}}\xspace}
\newcommand{\resulSetAKNNInEq}{\hyperref[tab:notations]{O_{aknn}}\xspace}
\newcommand{\recallInEq}{\hyperref[tab:notations]{r}\xspace}

\newcommand{\vectorSetInEq}{\hyperref[tab:notations]{S}}
\newcommand{\vectorSubsetInEq}{\hyperref[tab:notations]{C}}
\newcommand{\queryCommand}{\hyperref[tab:notations]{$Q$}\xspace}
\newcommand{\annk}{\hyperref[tab:notations]{k}\xspace}

\newcommand{\algoLSH}{\hyperref[tab:algosList]{\textsc{Lsh}}\xspace}
\newcommand{\algoPQ}{\hyperref[tab:algosList]{\textsc{Pq}}\xspace}
\newcommand{\algoIVFPQ}{\hyperref[tab:algosList]{\textsc{Ivfpq}}\xspace}
\newcommand{\algoOnlinePQ}{\hyperref[tab:algosList]{\textsc{onlinePQ}}\xspace}

\newcommand{\algoFlann}{\hyperref[tab:algosList]{\textsc{Flann}}\xspace}

\newcommand{\algoNSW}{\hyperref[tab:algosList]{\textsc{Nsw}}\xspace}
\newcommand{\algoNSG}{\hyperref[tab:algosList]{\textsc{Nsg}}\xspace}
\newcommand{\algoDPG}{\hyperref[tab:algosList]{\textsc{Dpg}}\xspace}
\newcommand{\algoNNDecent}{\hyperref[tab:algosList]{\textsc{Nndecent}}\xspace}
\newcommand{\algoHNSW}{\hyperref[tab:algosList]{\textsc{Hnsw}}\xspace}

\newcommand{\algoBF}{\hyperref[tab:algosList]{\textsc{Baseline}}\xspace}
\newcommand{\algoLSHAPG}{\hyperref[tab:algosList]{\textsc{Lshapg}}\xspace}
\newcommand{\DPR}{\hyperref[tab:realworld_workloads]{\textit{DPR}}\xspace}
\newcommand{\SIFT}{\hyperref[tab:realworld_workloads]{\textit{SIFT}}\xspace}
\newcommand{\Trevi}{\hyperref[tab:realworld_workloads]{\textit{Trevi}}\xspace}
\newcommand{\Glove}{\hyperref[tab:realworld_workloads]{\textit{Glove}}\xspace}
\newcommand{\Msong}{\hyperref[tab:realworld_workloads]{\textit{Msong}}\xspace}
\newcommand{\Sun}{\hyperref[tab:realworld_workloads]{\textit{Sun}}\xspace}
\newcommand{\Wte}{\hyperref[tab:realworld_workloads]{\textit{WTE}}\xspace}
\newcommand{\Random}{\hyperref[tab:realworld_workloads]{\textit{Random}}\xspace}

\newcommand{\compact}{\vspace{-4pt}}

%% file: authors.tex

\author{%
  Xianzhi Zeng$^1$, Zhuoyan Wu$^2$, Xinjing Hu$^3$, Xuanhua Shi$^3$, Shixuan Sun$^4$, Shuhao Zhang$^5$ \\  
  $^1$Singapore University of Technology and Design, $^2$National University of Singapore,\\ $^3$Huazhong University of Science and Technology, $^4$Shanghai Jiao Tong University,\\ $^5$Nanyang Technological University\\
  \texttt{xianzhi\_xianzhi@mymail.sutd.edu.sg}, \texttt{zhuoyan@comp.nus.edu.sg},\\ 
  \texttt{u202013895@hust.edu.cn},
  \texttt{xhshi@hust.edu.cn},\\
  \texttt{sunshixuan@sjtu.edu.cn},
  \texttt{shuhao.zhang@ntu.edu.sg} \\
}

%% file: abstract.tex
\begin{abstract}
Approximate K Nearest Neighbor (AKNN) algorithms play a pivotal role in various AI applications, including information retrieval, computer vision, and natural language processing. Although numerous AKNN algorithms and benchmarks have been developed recently to evaluate their effectiveness, the dynamic nature of real-world data presents significant challenges that existing benchmarks fail to address. Traditional benchmarks primarily assess retrieval effectiveness in static contexts and often overlook update efficiency, which is crucial for handling continuous data ingestion. This limitation results in an incomplete assessment of an AKNN algorithm's ability to adapt to changing data patterns, thereby restricting insights into their performance in dynamic environments.
To address these gaps, we introduce \system, a benchmark tailored for \underline{\textbf{C}}ontinuous \underline{\textbf{A}}pproximate \underline{\textbf{N}}earest Neighbor Search with \underline{\textbf{Dy}}namic Data Ingestion. \system comprehensively assesses a wide range of AKNN algorithms, integrating advanced optimizations such as machine learning-driven inference to supplant traditional heuristic scans, and improved distance computation methods to reduce computational overhead. Our extensive evaluations across diverse datasets demonstrate that simpler AKNN baselines often surpass more complex alternatives in terms of recall and latency. These findings challenge established beliefs about the necessity of algorithmic complexity for high performance. Furthermore, our results underscore existing challenges and illuminate future research opportunities.
We have made the datasets and implementation methods available at: \url{https://github.com/intellistream/candy}.
\end{abstract}

%% file: Introduction.tex
\section{Introduction}
\label{sec:intro}
Approximate K Nearest Neighbor (AKNN) search algorithms underpin many AI applications, including information retrieval~\cite{manning2008introduction}, computer vision~\cite{dong2003concept}, and natural language processing~\cite{luo2022semantic}. These algorithms support essential functions like real-time decision-making, image recognition, and semantic understanding~\cite{karpukhin2020dense, lewis2020retrieval}. AKNN algorithms have evolved significantly since their initial proposal in 1998~\cite{indyk1998approximate}. Early developments focused on optimizing search efficiency and accuracy through various techniques such as locality-sensitive hashing (LSH)\cite{gionis1999similarity}, tree-based indexing\cite{weber1998quantitative}, and graph-based methods~\cite{malkov2018efficient}, primarily for static datasets. Despite their importance, the evolving nature of real-world data presents significant challenges not adequately addressed by current benchmarks, which predominantly assess AKNN algorithms in static environments, focusing on retrieval effectiveness while largely ignoring update efficiency critical for ongoing data ingestion~\cite{li2019approximate, wang2021comprehensive, li2022deep}. This oversight results in an incomplete understanding of an AKNN algorithm's adaptability and performance in environments where it must continuously update and perform searches amid dynamic data ingestion.


Recognizing the limitations of traditional AKNN algorithms, there has been a recent shift towards adapting these methods for more realistic, continuously evolving data streams. The foundational work by Sundaram et al.\cite{sundaram2013streaming} first explored dynamic data ingestion in AKNN, investigating the incremental nature of Locality-Sensitive Hashing (\algoLSH\cite{gionis1999similarity}). Following this, \algoOnlinePQ~\cite{xu2018online} introduced incremental clustering to Product Quantization (\algoPQ~\cite{jegou2010product}) to dynamically adjust \algoPQ data structures as new information arrives. Researchers have also explored accommodating proxy graphs of AKNN, such as the Hierarchical Navigable Small Worlds (\algoHNSW~\cite{malkov2018efficient}), for incremental vector additions. Notably, the SSD-aware framework “Freshdiskann”\cite{singh2021freshdiskann} investigates vector compression and SSD-aware optimizations to support incremental vector additions. Recent developments by Aguerrebere et al.\cite{aguerrebere2024locally} further enhance AKNN algorithms with vector compression optimizations, marking significant progress in adapting AKNN algorithms to dynamic data ingestion.

However, the current literature~\cite{li2019approximate,wang2021comprehensive,li2022deep,karpukhin2020dense} still largely overlooks comprehensive methodologies for evaluating the efficiency and adaptability of AKNN algorithms in environments characterized by \textbf{continuous updates} and \textbf{distribution shifts}~\cite{aguerrebere2024locally,singh2021freshdiskann} due to dynamic data ingestion. This gap is particularly evident in scenarios like online news aggregators, where the need to adapt content recommendations in response to new articles and shifting user interactions highlights the inefficiencies of AKNN algorithms in updating indices with fresh data, potentially leading to stale recommendations and degraded user experiences. Such dynamics underscore a critical gap in existing benchmarking approaches and highlight the urgent need for new benchmarks that rigorously evaluate the resilience and adaptability of AKNN algorithms to meet the demands of modern, real-time digital ecosystems.

\begin{figure}[t]
    \centering
    \includegraphics[scale=0.4, bb=0 0 850 110]{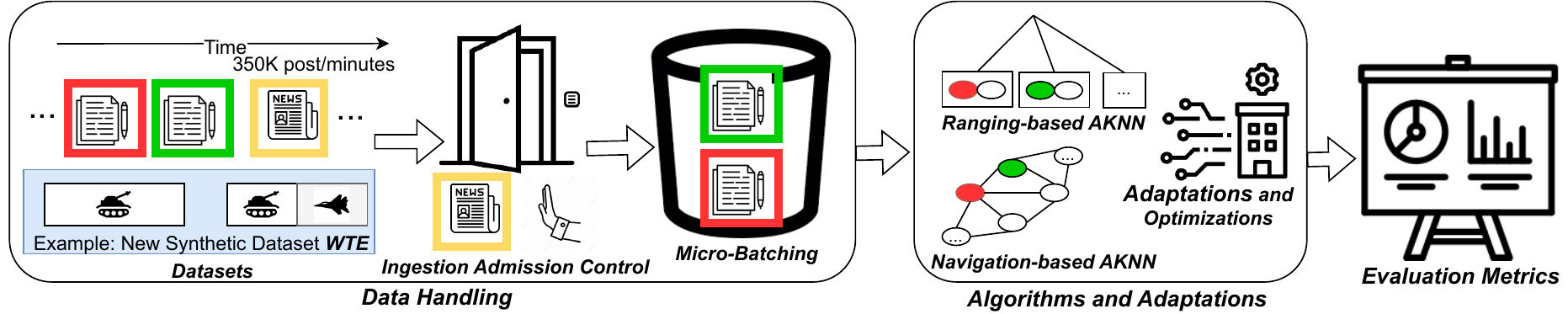}
    \caption{Overview of the \system benchmark framework.}
    \label{fig:overview}
\end{figure}

\textbf{Our Contributions:}
In response to the pressing need for benchmarks that effectively evaluate AKNN algorithms in dynamic data ingestion contexts, we have developed \system, a comprehensive benchmark framework (Figure~\ref{fig:overview}). Our contributions are articulated across several dimensions:

\begin{itemize}
    \item We introduce a novel categorization of state-of-the-art AKNN algorithms into two principal types: ranging-based and navigation-based. This classification facilitates a deeper understanding of the operational dynamics of these algorithms.
    \item We propose specific adaptations to improve the performance of AKNN algorithms, enabling them to continuously adapt to dynamic data ingestion. We have also implemented optimizations that significantly boost both the efficiency and accuracy of these algorithms across various data conditions.
    \item \system incorporates advanced data handling strategies such as data dropping and micro-batching, which simulate real-world dynamic data ingestion. This setup facilitates rigorous testing of AKNN algorithms under realistic rates and conditions. Furthermore, to explore the effects of distribution shifts, we have developed a novel synthetic dataset \Wte.
    \item Utilizing \system, our experimental studies provide novel insights into the adaptability and overall performance of AKNN under the aforementioned common concerns, i.e.,
    the \emph{continuous updates} and \emph{distribution shifts}.
    We further examine the effectiveness of \emph{AKNN optimizations} and other important configurations such as {micro-batching}.
    
\end{itemize}

%% file: BenchmarkSystem.tex
\section{\system Benchmark Framework}
\label{sec:benchmark_framework}
We now present \system, a benchmark for evaluating continuous AKNN with dynamic data ingestion.

\input{BenchmarkSystems/DataHandling}
\input{BenchmarkSystems/Algos}
\input{BenchmarkSystems/Metrics}


%% file: BenchmarkSystems/DataHandling.tex
\subsection{Data Handling}
\label{subsec:data_handling}


\paragraph{Datasets}
Consistent with methodologies from previous research~\cite{li2019approximate, aguerrebere2024locally}, we utilize a variety of real-world workloads spanning multiple application domains, including text embedding (\Glove, \DPR), image processing (\SIFT, \Sun, \Trevi), and audio (\Msong). We also incorporate the \Random synthetic dataset from LibTorch~\cite{Pytorch}, which offers adjustable dimensions and volumes of vectors and employs the \emph{torch::rand} function to generate identically distributed (i.i.d.) uniform values between $0$ and $1$. Recognizing that the performance of AKNN algorithms can be markedly influenced by drifts in data distribution~\cite{aguerrebere2024locally}, such as evolving news topics, we have developed the \Wte dataset to effectively simulate these dynamic changes. Derived from nouns categorized as land vehicles and aircraft in the War Thunder game wiki~\cite{warthunderWiki}, \Wte uses these nouns to create synthetic sentences, subsequently encoded into vectors via the \DPR methodology~\cite{aguerrebere2024locally}. This dataset allows for the adjustment of drift starting positions by altering the onset of new noun categories and enables the modulation of drift intensities through controlled word contamination, with the magnitude dictated by predefined probabilities. More details can be found at Appendix~\ref{subsec:detail_benchmark}.

\paragraph{Ingestion Admission Control} The rapid influx of data, such as high-rate tweets~\cite{sundaram2013streaming,tweetsSpeed}, imposes significant time and computational constraints on processing capabilities. These constraints, often overlooked in AKNN literature, critically hinder the system's ability to maintain an up-to-date index. If the processing speed of AKNN algorithms lags behind the incoming data flow, an unmanageable backlog may develop, potentially growing indefinitely~\cite{cidon2015tiered,andrews2004scheduling,fowler1991local,arun2018copa}. Pausing data input to clear the backlog is impractical due to limited storage for pending data, forcing the system to discard incoming information. This often results in decreased accuracy and reduced search efficiency. To accurately simulate this challenge, \system includes a specific ingestion admission control protocol. This protocol utilizes a high-resolution clock to regulate data release at a constant rate, typically set at 4,000 rows per second by default. If AKNN processes data more slowly than this rate, excess data is naturally dropped due to congestion, mirroring realistic operational constraints.

\paragraph{Micro-Batching}
The data science community recognizes the significant advantages of micro-batching for effectively processing large datasets, though the specific motivations and purposes for its use can vary. Micro-batching divides extensive datasets into smaller, manageable units, enabling more frequent updates to machine learning model parameters and thus promoting quicker model convergence~\cite{huang2019gpipe,yuan2022decentralized}. On the other hand, accumulating data in larger batches instead of processing streams immediately can greatly improve the throughput of online database systems~\cite{zhang2021parallelizing}. Inspired by these insights, we have integrated adaptable micro-batch mechanisms into \system, allowing for the aggregation of continuously ingested data streams into batches with configurable sizes.


%% file: BenchmarkSystems/Algos.tex
\subsection{Categorization, Adaptation, and Optimization}
\label{subsec:aknn_algorithm_categorization}

AKNN aims to efficiently identify the \annk \emph{closest} vectors to a query \queryCommand from a set, \vectorSet. AKNN algorithms optimize search processes by selecting a strategic subset of vectors, \vectorSubset (i.e., $\vectorSubsetInEq \subseteq \vectorSetInEq$), which significantly enhances data retrieval efficiency compared to exhaustive search baseline (denoted as \algoBF). 
Common distance measures include L2 distance~\cite{li2019approximate} and inner product distance~\cite{aguerrebere2024locally}, and both are implemented in \system and inner product is used unless mentioned otherwise.

\paragraph{Algorithm Categorization} 
We classify AKNN algorithms into two main categories based on their subset selection techniques: \emph{ranging-based} and \emph{navigation-based}. This classification helps assess how different AKNN strategies manage the trade-offs between computational overhead and accuracy. The performance and adaptability of each category are thoroughly evaluated and summarized in Table~\ref{tab:algosList}, providing insights into their effectiveness in handling real-time data variations.

\emph{1) Ranging-based AKNN:} Ranging-based AKNN involves pre-computed division of \vectorSet into distinct ranges, irrespective of the specific \queryCommand. It segments \vectorSet into disjoint ranges such as $D_1, D_2, \ldots, D_n$. Upon receiving a \queryCommand, \vectorSubset is selected from one or a combination of these ranges (e.g., $\vectorSubsetInEq = D_1$ or $\vectorSubsetInEq = D_1 \cup D_2$). \system includes a spectrum of ranging-based AKNN algorithms, from traditional approaches like \algoLSH to those specially designed for dynamic data scenarios, such as \algoOnlinePQ.

\emph{2) Navigation-based AKNN:} 
Navigation-based AKNN establishes navigational links within \vectorSet, associating each vector with its similar counterparts, typically through a proximity graph. Unlike the static pre-computed ranges of the ranging-based approach, the subset \vectorSubset for each query is dynamically assembled by iteratively navigating from one vector to its neighboring vectors. Key variations among navigation-based AKNN methods include their strategies for navigating and deciding when to cease expanding the pool of candidate neighbors. Prominent implementations in \system feature the K-Nearest Neighbor Graph (\algoNNDecent) and hierarchical approximate Delaunay Graph (\algoHNSW).

\begin{table}[]
\centering
\resizebox{0.8\textwidth}{!}{%
\begin{tabular}{|c|l|l|}
\hline
\multicolumn{1}{|c|}{\textit{Category}} & \textit{Algorithm Name}                & \textit{Descriptions}                                                                                             \\ \hline
\multirow{6}{*}{Ranging-based}        
& \algoLSH~\cite{gionis1999similarity}   & Data-independednt hashing to determine ranges                                                                     \\ \cline{2-3} 
 & \algoFlann~\cite{muja2014scalable}     & Space partition to determine ranges                                                                               \\ \cline{2-3} 
                                        & \algoPQ~\cite{jegou2010product}        & Product quantization to determin ranges                                                                           \\ \cline{2-3} 
                                        & \algoIVFPQ~\cite{jegou2010product}     & Hierachical optimization over \algoPQ                                                                             \\ \cline{2-3} 
                                       
                                        & \algoOnlinePQ~\cite{xu2018online}        & Incrementally update centroids of \algoPQ 
                                                                                  \\ \hline
\multirow{6}{*}{Navigation-based}       & \algoNSW~\cite{malkov2014approximate}  & Delaunay Graph for navigation                                                                                     \\ \cline{2-3} 
                                        & \algoNNDecent~\cite{dong2011efficient} & K-Nearest Neighbor Graph for navigation                                                                           \\ \cline{2-3} 
                                        & \algoDPG~\cite{li2019approximate}      & \begin{tabular}[c]{@{}l@{}}K-Nearest Neighbor Graph and Relative Neighborhood Graph\\ for navigation\end{tabular} \\ \cline{2-3} 
                                        & \algoNSG~\cite{fu2019fast}      & \begin{tabular}[c]{@{}l@{}} Optimized vertex assignment compared with \algoDPG \end{tabular} \\ \cline{2-3} 
                                        & \algoHNSW~\cite{malkov2018efficient}   & Hiearchical construction of \algoNSW     \\ \cline{2-3}          
                                        & \algoLSHAPG~\cite{zhao2023towards} & Using the hashing of \algoLSH to optimize graph construction
                                        \\ \hline
\end{tabular}
}
\caption{Representative AKNN algorithms and their categories.}
\label{tab:algosList}
\end{table}

\paragraph{Adapting to Dynamic Data Ingestion}
We discuss the adaptation of AKNN algorithms to dynamic data ingestion, highlighting three distinct approaches:

\emph{1) Minimal Adaptation Required:} 
Algorithms such as \algoLSH, \algoNSW, \algoHNSW and \algoLSHAPG inherently support incremental updates and are thus naturally suited for dynamic data environments. \algoOnlinePQ, designed explicitly for dynamic ingestion, fits seamlessly with minimal modifications needed. For these algorithms, adaptation primarily involves aligning their API and I/O structures within our codebase to ensure compatibility.

\emph{2) Adaptation with Enforced Assumptions:} 
\algoPQ and \algoIVFPQ process data incrementally but depend on a pre-existing understanding of data distributions, typically acquired through extensive preliminary data scanning. This process, while intensive, is necessary to set up the initial knowledge base used during dynamic ingestion. In our framework, we establish this knowledge base during an offline stage and maintain the assumption that it remains valid in dynamic contexts. We exclude any overhead from this setup phase in our performance evaluations to focus on real-time processing capabilities.

\emph{3) Substantial Algorithm Modifications:}
Algorithms like \algoNNDecent, \algoDPG, and particularly \algoNSG, initially designed for static datasets, require comprehensive modifications to accommodate dynamic data ingestion. \algoNNDecent originally constructs a neighborhood graph offline, with each node maintaining its closest k nodes in a heap. For dynamic adaptation, we have integrated online query and update capabilities. The online query algorithm initiates from a randomly selected node, employing a greedy strategy to converge towards the target region while maintaining a heap of candidate nodes for iterative expansion. Online insertion involves identifying and updating the neighbors of the newly inserted node through this online search mechanism, effectively embedding the node into the graph. Online deletions are managed by marking nodes in a deleted set, thus excluding them from query results without physically removing them from the graph.

\algoDPG builds upon the neighborhood graph to create layered structures. It modifies the initial graph (layer0) by selecting only half of each node's neighbors based on directional diversity rather than distance, aiming to disperse neighbors across various directions and maintaining bidirectional connectivity. The query process mirrors that of \algoNNDecent, targeting the modified graph (layer1). The online insertion updates both layers: it first updates layer0 with new neighbors as identified by the online search and then adjusts layer1 neighbors for the newly inserted node using a \(O(K^2)\) selection algorithm. If a node becomes a neighbor in layer0, it is also updated in layer1, either by direct addition if there is capacity or by replacing an existing neighbor if the displaced relationship is also reflected in layer1.

\algoNSG uses a pre-built graph based on brute force. Adapting \algoNSG for dynamic use involves integrating newly ingested data points into the existing graph structure. Specifically, each new point undergoes a global search for neighbors starting from a designated navigation point, similar to the initial build phase. This process ensures that the graph remains navigable and that new points are appropriately linked to maintain the integrity and efficiency of the graph, allowing the algorithm to continue to function effectively as new data is ingested.

\paragraph{Optimization Techniques}
\label{subsub:aknn_opts}
Recent AKNN optimization techniques are diverse but can be strategically categorized into two main types: machine learning and optimized distance computation~\cite{li2023learning,NIPS2008_d58072be,baranchuk2019learning,aguerrebere2024locally,gao2023high}. Each technique is designed for specific AKNN algorithms or distance metrics, and both are integrated into \system.

\emph{1) Machine learning-based optimizations} are employed for both ranging-based~\cite{li2023learning,NIPS2008_d58072be} and navigation-based~\cite{baranchuk2019learning} AKNN algorithms. These optimizations typically replace costly heuristic scans or vector data traversals with more efficient machine learning inferences, transforming data access overhead into computationally cheaper mathematical calculations~\cite{li2022deep}. \system, leveraging the LibTorch/PyTorch ecosystem, supports standard neural network modules derived from \emph{torch::nn::Module}, tensor-based loss functions, and commonly used optimizers such as \emph{torch::optim::Adam}. This flexibility facilitates the exploration of advanced machine learning techniques to further enhance AKNN performance.

\emph{2) Distance computation optimization} is another crucial strategy~\cite{aguerrebere2024locally,gao2023high}. The objective is to minimize the computational load associated with calculating and comparing distances among vectors, a significant bottleneck in many AKNN applications. Techniques such as randomized calculations~\cite{gao2023high} and data compression~\cite{aguerrebere2024locally} are implemented to reduce these overheads effectively. Together, these optimization strategies provide a comprehensive approach to refining the efficiency and accuracy of AKNN algorithms within \system.


 

%% file: BenchmarkSystems/Metrics.tex

\subsection{Performance Evaluation Metrics}
Following established methodologies~\cite{li2019approximate,gao2023high,li2022deep}, we employ \emph{query recall} (\recall) to evaluate the accuracy of AKNN algorithms. Query recall measures the proportion of the true nearest neighbors that are correctly identified by the algorithm from all possible true nearest neighbors. Specifically, if the set of nearest neighbors identified by AKNN is \resulSetAKNN, and the exact set of nearest neighbors is \resulSet, then \recall is defined as $\recallInEq = \frac{|\resulSetAKNNInEq \cap \resulSetInEq|}{|\resulSetInEq|}$. By default, we report the recall for retrieving the top 10 nearest vectors for each of 100 input queries, denoted as \emph{recall@10} in following~\cite{aguerrebere2024locally}.
Additionally, we assess the efficiency of AKNN algorithms using \emph{query latency}, which measures the elapsed time from the submission of a query to when AKNN returns the results~\cite{li2019approximate,gao2023high,li2022deep}. Importantly, we exclude any setup overhead incurred during offline stages, such as loading initial data or preparing data structures, from this metric. This exclusion ensures that our latency measurements reflect only the real-time processing capabilities of the AKNN algorithms.

%% file: ExperimentShort.tex
\subsection{Summary of Differences from Static Data Scenarios}
\label{subsub:differ}
Our research deviates from traditional AKNN evaluations, which typically preload all data during an offline stage without subsequent updates to the data storage. Unlike these conventional setups referenced in prior studies~\cite{li2019approximate,gionis1999similarity,jegou2010product,muja2014scalable,xu2018online,malkov2014approximate,li2019approximate,fu2019fast}, we examine the impact of dynamic data ingestion on AKNN performance. Our methodology starts with an initial dataset loaded into AKNN, then transitions to a phase of continuous online data feeding, mimicking real-world scenarios where data is not static but constantly arriving. This setup allows us to explore operational challenges unique to dynamic environments, such as data overflow and the processing demands of simultaneous data ingestion and query handling, which can degrade query recall and increase latency.

\section{Experiments and Observations}
\label{sec:insights}
In this section, we detail our experimental approach using the \system framework to assess how AKNN algorithms perform when subjected to continuous data flow.
Note that, we have adhered to recommended default settings from prominent studies~\cite{gionis1999similarity,jegou2010product,muja2014scalable,xu2018online,malkov2014approximate,li2019approximate,fu2019fast,zhao2023towards}, validated to balance accuracy and efficiency, without specifically tuning any algorithm to maximize recall or minimize latency~\cite{li2019approximate}.




\newcounter{myvar}
\setcounter{myvar}{1}

\input{EvaluationShort/Overall}

\input{EvaluationShort/Shift}
\input{EvaluationShort/Agglomeration}

\input{EvaluationShort/Opt}
\label{sec:methodology}

%% file: EvaluationShort/Overall.tex
\subsection{Poor Ingestion Efficiency is a Severe Common Bottleneck}
\label{subsec:overall}
We first investigate how different AKNN algorithms performs under real-world datasets, concerning both query latency and query recall, as summarized in Table~\ref{tab:e2e_ip}. 
\begin{quote}
    Observation \themyvar. Contrary to common belief~\cite{li2019approximate}, none of the evaluated AKNN  outperforms the brutal force approach with dynamic data ingestion. This is majorly due to the large overhead of pending writings in AKNN.
\end{quote}
\stepcounter{myvar}
\begin{table}[]
\centering
 \resizebox{0.9\textwidth}{!}{%
\begin{tabular}{|ll|llllll|llllll|}
\hline
\multicolumn{2}{|l|}{\multirow{2}{*}{Algorithms}}                                & \multicolumn{6}{l|}{Recall@10}                                                                                                                                                                         & \multicolumn{6}{l|}{Query Latency ($\times 1000 ms$)}                                                                                                                                                               \\ \cline{3-14} 
\multicolumn{2}{|l|}{}                                                           & \multicolumn{1}{l|}{\Glove}        & \multicolumn{1}{l|}{\SIFT}         & \multicolumn{1}{l|}{\Msong}        & \multicolumn{1}{l|}{\Sun}          & \multicolumn{1}{l|}{\DPR}          & \Trevi        & \multicolumn{1}{l|}{\Glove}          & \multicolumn{1}{l|}{\SIFT}          & \multicolumn{1}{l|}{\Msong}          & \multicolumn{1}{l|}{\Sun}            & \multicolumn{1}{l|}{\DPR}            & \Trevi            \\ \hline
\multicolumn{2}{|l|}{\textbf{\algoBF}}                                           & \multicolumn{1}{l|}{\textbf{1.00}} & \multicolumn{1}{l|}{\textbf{1.00}} & \multicolumn{1}{l|}{\textbf{1.00}} & \multicolumn{1}{l|}{\textbf{1.00}} & \multicolumn{1}{l|}{\textbf{1.00}} & \textbf{1.00} & \multicolumn{1}{l|}{\textbf{0.43}}   & \multicolumn{1}{l|}{\textbf{0.44}}  & \multicolumn{1}{l|}{\textbf{0.60}}   & \multicolumn{1}{l|}{\textbf{0.69}}   & \multicolumn{1}{l|}{\textbf{0.76}}   & \textbf{2.53}     \\ \hline
\multicolumn{1}{|l|}{\multirow{5}{*}{Ranging-based}}    & \textbf{\algoLSH}      & \multicolumn{1}{l|}{\textbf{0.00}} & \multicolumn{1}{l|}{\textbf{0.01}} & \multicolumn{1}{l|}{\textbf{0.00}} & \multicolumn{1}{l|}{\textbf{0.00}} & \multicolumn{1}{l|}{\textbf{0.00}} & \textbf{0.07} & \multicolumn{1}{l|}{\textbf{0.17}}   & \multicolumn{1}{l|}{\textbf{0.22}}  & \multicolumn{1}{l|}{\textbf{0.23}}   & \multicolumn{1}{l|}{\textbf{0.42}}   & \multicolumn{1}{l|}{\textbf{0.19}}   & \textbf{11.96}    \\ \cline{2-14} 
\multicolumn{1}{|l|}{}                                  & \algoFlann             & \multicolumn{1}{l|}{0.01}          & \multicolumn{1}{l|}{0.01}          & \multicolumn{1}{l|}{0.01}          & \multicolumn{1}{l|}{0.00}          & \multicolumn{1}{l|}{0.00}          & 0.11          & \multicolumn{1}{l|}{1.09}            & \multicolumn{1}{l|}{1.18}           & \multicolumn{1}{l|}{1.85}            & \multicolumn{1}{l|}{1.65}            & \multicolumn{1}{l|}{1.27}            & 2.69              \\ \cline{2-14} 
\multicolumn{1}{|l|}{}                                  & \algoPQ                & \multicolumn{1}{l|}{0.09}          & \multicolumn{1}{l|}{0.22}          & \multicolumn{1}{l|}{0.50}          & \multicolumn{1}{l|}{0.42}          & \multicolumn{1}{l|}{0.09}          & 0.38          & \multicolumn{1}{l|}{354.43}          & \multicolumn{1}{l|}{351.58}         & \multicolumn{1}{l|}{358.40}          & \multicolumn{1}{l|}{363.92}          & \multicolumn{1}{l|}{361.51}          & 414.77            \\ \cline{2-14} 
\multicolumn{1}{|l|}{}                                  & \algoIVFPQ             & \multicolumn{1}{l|}{0.03}          & \multicolumn{1}{l|}{0.23}          & \multicolumn{1}{l|}{0.01}          & \multicolumn{1}{l|}{0.33}          & \multicolumn{1}{l|}{0.11}          & 0.28          & \multicolumn{1}{l|}{393.40}          & \multicolumn{1}{l|}{2.11}           & \multicolumn{1}{l|}{400.55}          & \multicolumn{1}{l|}{2.85}            & \multicolumn{1}{l|}{409.62}          & 504.46            \\ \cline{2-14} 
\multicolumn{1}{|l|}{}                                  & \textbf{\algoOnlinePQ} & \multicolumn{1}{l|}{\textbf{0.12}} & \multicolumn{1}{l|}{\textbf{0.31}} & \multicolumn{1}{l|}{\textbf{0.78}} & \multicolumn{1}{l|}{\textbf{0.43}} & \multicolumn{1}{l|}{\textbf{0.09}} & \textbf{0.38} & \multicolumn{1}{l|}{\textbf{1.60}}   & \multicolumn{1}{l|}{\textbf{1.20}}  & \multicolumn{1}{l|}{\textbf{3.36}}   & \multicolumn{1}{l|}{\textbf{384.48}} & \multicolumn{1}{l|}{\textbf{395.55}} & \textbf{19.20}    \\ \hline
\multicolumn{1}{|l|}{\multirow{6}{*}{Navigation-based}} & \textbf{\algoHNSW}     & \multicolumn{1}{l|}{\textbf{0.39}} & \multicolumn{1}{l|}{\textbf{0.54}} & \multicolumn{1}{l|}{\textbf{0.61}} & \multicolumn{1}{l|}{\textbf{0.77}} & \multicolumn{1}{l|}{\textbf{0.46}} & \textbf{0.51} & \multicolumn{1}{l|}{\textbf{117.76}} & \multicolumn{1}{l|}{\textbf{15.86}} & \multicolumn{1}{l|}{\textbf{151.58}} & \multicolumn{1}{l|}{\textbf{584.71}} & \multicolumn{1}{l|}{\textbf{691.58}} & \textbf{13466.48} \\ \cline{2-14} 
\multicolumn{1}{|l|}{}                                  & \algoNSW               & \multicolumn{1}{l|}{0.10}          & \multicolumn{1}{l|}{0.35}          & \multicolumn{1}{l|}{0.28}          & \multicolumn{1}{l|}{0.63}          & \multicolumn{1}{l|}{0.33}          & 0.51          & \multicolumn{1}{l|}{1883.68}         & \multicolumn{1}{l|}{233.17}         & \multicolumn{1}{l|}{454.67}          & \multicolumn{1}{l|}{770.08}          & \multicolumn{1}{l|}{1190.43}         & 7048.91           \\ \cline{2-14} 
\multicolumn{1}{|l|}{}                                  & \algoNSG               & \multicolumn{1}{l|}{0.01}          & \multicolumn{1}{l|}{0.01}          & \multicolumn{1}{l|}{0.00}          & \multicolumn{1}{l|}{0.02}          & \multicolumn{1}{l|}{0.01}          & 0.03          & \multicolumn{1}{l|}{30.62}           & \multicolumn{1}{l|}{323.19}         & \multicolumn{1}{l|}{401.75}          & \multicolumn{1}{l|}{46.20}           & \multicolumn{1}{l|}{48.51}           & 1262.32           \\ \cline{2-14} 
\multicolumn{1}{|l|}{}                                  & \algoNNDecent          & \multicolumn{1}{l|}{0.01}          & \multicolumn{1}{l|}{0.12}          & \multicolumn{1}{l|}{0.00}          & \multicolumn{1}{l|}{0.23}          & \multicolumn{1}{l|}{0.17}          & 0.46          & \multicolumn{1}{l|}{30.58}           & \multicolumn{1}{l|}{15.38}          & \multicolumn{1}{l|}{5.34}            & \multicolumn{1}{l|}{6.61}            & \multicolumn{1}{l|}{40.38}           & 20.23             \\ \cline{2-14} 
\multicolumn{1}{|l|}{}                                  & \algoDPG               & \multicolumn{1}{l|}{0.08}          & \multicolumn{1}{l|}{0.33}          & \multicolumn{1}{l|}{0.06}          & \multicolumn{1}{l|}{0.30}          & \multicolumn{1}{l|}{0.31}          & 0.48          & \multicolumn{1}{l|}{18.81}           & \multicolumn{1}{l|}{19.79}          & \multicolumn{1}{l|}{614.12}          & \multicolumn{1}{l|}{692.44}          & \multicolumn{1}{l|}{87.52}           & 5547.55           \\ \cline{2-14} 
\multicolumn{1}{|l|}{}                                  & \algoLSHAPG            & \multicolumn{1}{l|}{0.11}          & \multicolumn{1}{l|}{0.10}          & \multicolumn{1}{l|}{0.49}          & \multicolumn{1}{l|}{0.30}          & \multicolumn{1}{l|}{0.11}          & 0.42          & \multicolumn{1}{l|}{11.52}           & \multicolumn{1}{l|}{4.72}           & \multicolumn{1}{l|}{20.20}           & \multicolumn{1}{l|}{27.90}           & \multicolumn{1}{l|}{8.14}            & 22.37             \\ \hline
\end{tabular}
}
\caption{Comparing AKNN algorithms with dynamic data ingestion.}
\label{tab:e2e_ip}
\end{table}

We observe that \algoLSH surpasses \algoBF in terms of query latency on certain datasets, achieving up to 75\% reduction on \DPR. However, its accuracy is significantly lower, with a maximum recall of only 0.01, rendering it too imprecise for practical application. In contrast, \algoOnlinePQ and \algoHNSW demonstrate the highest accuracy among ranging-based and navigation-based AKNN algorithms, respectively, with recalls reaching as high as 0.7. Despite their superior accuracy, these algorithms exhibit considerably higher query latencies compared to \algoBF; for instance, \algoHNSW's latency is 908 times greater on \DPR.

\begin{table}[]
\centering
 \resizebox{0.9\textwidth}{!}{%
\begin{tabular}{|ll|lll|lll|lll|}
\hline
\multicolumn{2}{|l|}{\multirow{2}{*}{Algorithms}}                                & \multicolumn{3}{l|}{Propotion of PWL (\%)}                                                 & \multicolumn{3}{l|}{PWL ($\times 1000ms$)}                                                      & \multicolumn{3}{l|}{VSL ($\times 1000ms$)}                                                  \\ \cline{3-11} 
\multicolumn{2}{|l|}{}                                                           & \multicolumn{1}{l|}{\Msong}         & \multicolumn{1}{l|}{\DPR}           & \Trevi         & \multicolumn{1}{l|}{\Msong}          & \multicolumn{1}{l|}{\DPR}            & \Trevi            & \multicolumn{1}{l|}{\Msong}        & \multicolumn{1}{l|}{\DPR}           & \Trevi           \\ \hline
\multicolumn{2}{|l|}{\textbf{\algoBF}}                                           & \multicolumn{1}{l|}{\textbf{73.13}} & \multicolumn{1}{l|}{\textbf{63.32}} & \textbf{59.96} & \multicolumn{1}{l|}{\textbf{0.44}}   & \multicolumn{1}{l|}{\textbf{0.48}}   & \textbf{1.51}     & \multicolumn{1}{l|}{\textbf{0.16}} & \multicolumn{1}{l|}{\textbf{0.28}}  & \textbf{1.01}    \\ \hline
\multicolumn{1}{|l|}{\multirow{5}{*}{Ranging-based}}    & \algoLSH               & \multicolumn{1}{l|}{74.99}          & \multicolumn{1}{l|}{89.10}          & 1.11           & \multicolumn{1}{l|}{0.17}            & \multicolumn{1}{l|}{0.17}            & 0.13              & \multicolumn{1}{l|}{0.06}          & \multicolumn{1}{l|}{0.02}           & 11.83            \\ \cline{2-11} 
\multicolumn{1}{|l|}{}                                  & \algoFlann             & \multicolumn{1}{l|}{98.99}          & \multicolumn{1}{l|}{98.14}          & 97.48          & \multicolumn{1}{l|}{1.83}            & \multicolumn{1}{l|}{1.24}            & 2.62              & \multicolumn{1}{l|}{0.02}          & \multicolumn{1}{l|}{0.02}           & 0.07             \\ \cline{2-11} 
\multicolumn{1}{|l|}{}                                  & \algoPQ                & \multicolumn{1}{l|}{99.55}          & \multicolumn{1}{l|}{99.61}          & 99.53          & \multicolumn{1}{l|}{356.80}          & \multicolumn{1}{l|}{360.10}          & 412.80            & \multicolumn{1}{l|}{1.60}          & \multicolumn{1}{l|}{1.40}           & 1.97             \\ \cline{2-11} 
\multicolumn{1}{|l|}{}                                  & \algoIVFPQ             & \multicolumn{1}{l|}{99.52}          & \multicolumn{1}{l|}{99.55}          & 99.16          & \multicolumn{1}{l|}{398.64}          & \multicolumn{1}{l|}{407.78}          & 500.24            & \multicolumn{1}{l|}{1.92}          & \multicolumn{1}{l|}{1.83}           & 4.22             \\ \cline{2-11} 
\multicolumn{1}{|l|}{}                                  & \textbf{\algoOnlinePQ} & \multicolumn{1}{l|}{\textbf{94.38}} & \multicolumn{1}{l|}{\textbf{99.14}} & \textbf{99.15} & \multicolumn{1}{l|}{\textbf{3.17}}   & \multicolumn{1}{l|}{\textbf{392.14}} & \textbf{19.03}    & \multicolumn{1}{l|}{\textbf{0.19}} & \multicolumn{1}{l|}{\textbf{3.41}}  & \textbf{0.16}    \\ \hline
\multicolumn{1}{|l|}{\multirow{6}{*}{Navigation-based}} & \textbf{\algoHNSW}     & \multicolumn{1}{l|}{\textbf{99.96}} & \multicolumn{1}{l|}{\textbf{88.94}} & \textbf{86.87} & \multicolumn{1}{l|}{\textbf{151.52}} & \multicolumn{1}{l|}{\textbf{615.08}} & \textbf{11698.58} & \multicolumn{1}{l|}{\textbf{0.06}} & \multicolumn{1}{l|}{\textbf{76.49}} & \textbf{1767.90} \\ \cline{2-11} 
\multicolumn{1}{|l|}{}                                  & \algoNSW               & \multicolumn{1}{l|}{99.98}          & \multicolumn{1}{l|}{90.48}          & 88.04          & \multicolumn{1}{l|}{454.59}          & \multicolumn{1}{l|}{1077.06}         & 6205.80           & \multicolumn{1}{l|}{0.09}          & \multicolumn{1}{l|}{113.37}         & 843.11           \\ \cline{2-11} 
\multicolumn{1}{|l|}{}                                  & \algoNSG               & \multicolumn{1}{l|}{94.33}          & \multicolumn{1}{l|}{99.94}          & 94.80          & \multicolumn{1}{l|}{378.95}          & \multicolumn{1}{l|}{48.48}           & 1196.68           & \multicolumn{1}{l|}{22.80}         & \multicolumn{1}{l|}{0.03}           & 65.63            \\ \cline{2-11} 
\multicolumn{1}{|l|}{}                                  & \algoNNDecent          & \multicolumn{1}{l|}{98.77}          & \multicolumn{1}{l|}{98.83}          & 97.46          & \multicolumn{1}{l|}{5.27}            & \multicolumn{1}{l|}{39.90}           & 19.71             & \multicolumn{1}{l|}{0.07}          & \multicolumn{1}{l|}{0.47}           & 0.51             \\ \cline{2-11} 
\multicolumn{1}{|l|}{}                                  & \algoDPG               & \multicolumn{1}{l|}{98.26}          & \multicolumn{1}{l|}{99.33}          & 98.29          & \multicolumn{1}{l|}{603.45}          & \multicolumn{1}{l|}{86.93}           & 5452.70           & \multicolumn{1}{l|}{10.67}         & \multicolumn{1}{l|}{0.59}           & 94.85            \\ \cline{2-11} 
\multicolumn{1}{|l|}{}                                  & \algoLSHAPG            & \multicolumn{1}{l|}{80.96}          & \multicolumn{1}{l|}{93.15}          & 94.20          & \multicolumn{1}{l|}{16.36}           & \multicolumn{1}{l|}{7.58}            & 21.07             & \multicolumn{1}{l|}{3.85}          & \multicolumn{1}{l|}{0.56}           & 1.30             \\ \hline
\end{tabular}
}
\caption{Break down the query latency into pending write latency and vector search latency. ``PWL'' is short for pending writing latency, and ``VSL'' is short for vector search latency. }
\label{tab:e2e_bd}

\end{table}

In Table~\ref{tab:e2e_bd}, query latency is segmented into two distinct components for clearer analysis: \emph{pending writing latency}, which accounts for the time spent waiting for ongoing data writes to complete, and \emph{vector search latency}, which measures the time taken for vector search operations after AKNN has processed pending writes. \Msong, \DPR, and \Trevi are used as example datasets, with similar observations noted for other datasets, though these results are omitted.
Two primary insights emerge from this analysis. First, for the majority of AKNN algorithms across most datasets, pending writing latency constitutes the larger portion of total query latency. For example, in \algoOnlinePQ and \algoHNSW, pending writing accounts for over 94\% and 86\% of the total query latency, respectively. This indicates that pending writing is a significant bottleneck in the performance of query processing.
Second, while the vector search latency for some AKNN algorithms is lower than that of \algoBF in certain datasets, this advantage is not consistent across all datasets. For instance, in the \Msong dataset, \algoHNSW reduces vector search latency by 62\% compared to \algoBF, with a reasonable recall of 0.61, as shown in Table~\ref{tab:e2e_ip}. However, in the \DPR and \Trevi datasets, \algoHNSW exhibits vector search latencies that are $271\times$ and $1766\times$ longer than \algoBF, respectively. This inconsistency can be attributed to the dropping of critical data during ingestion, which complicates the navigation process and thereby increases the latency.

%% file: EvaluationShort/Shift.tex
\subsection{Most AKNN Acts Poorly Towards Distribution Drifts}
We next examine the impacts of distribution drifts, focusing specifically on the effects of drift occurrence position and general intensity of distribution drifts, which we analyze individually. Our evaluations are conducted using our novel datasets \Wte.

\begin{quote}
    Observation \themyvar.
    {Most AKNNs except \algoHNSW are challenged by distribution drifts. Despite the resilience, the ingestion bottleneck of \algoHNSW is affected by the distribution shift and further complicated by the semantic relevance of data.}
\end{quote}
\stepcounter{myvar}
\begin{figure}[t]
\begin{minipage}[a]{\textwidth} 

         \begin{minipage}[b] {0.70\textwidth}
        \begin{minipage}[c]{0.57\textwidth}
        \centering
        \subfigure[Query Recall]{
            \includegraphics[width=0.99\textwidth]{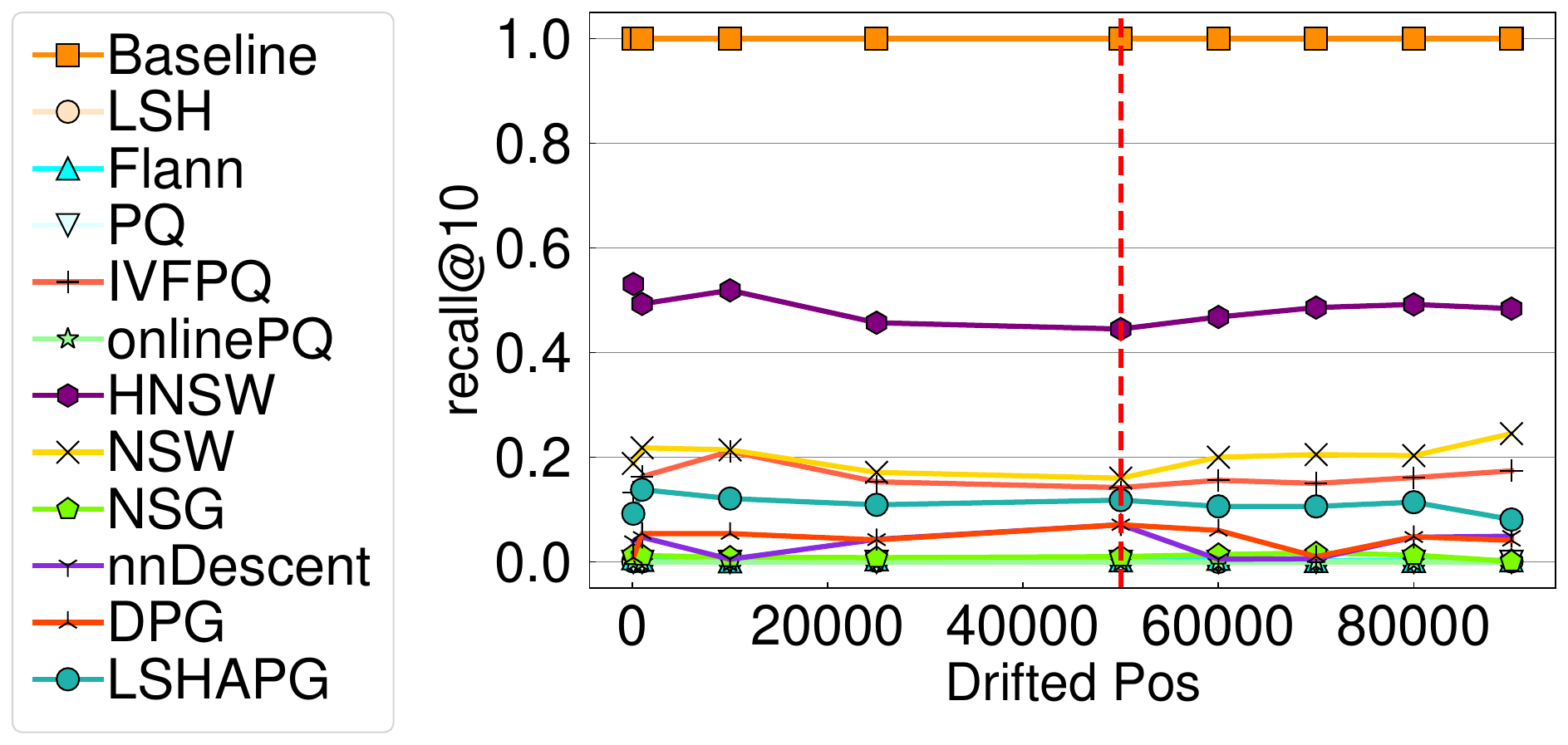}
            \label{fig:cd_recall}
        }
    \end{minipage}
    \begin{minipage}[c]{0.41\textwidth}
        \subfigure[Query Latency.]{
            \includegraphics[width=0.99\textwidth]{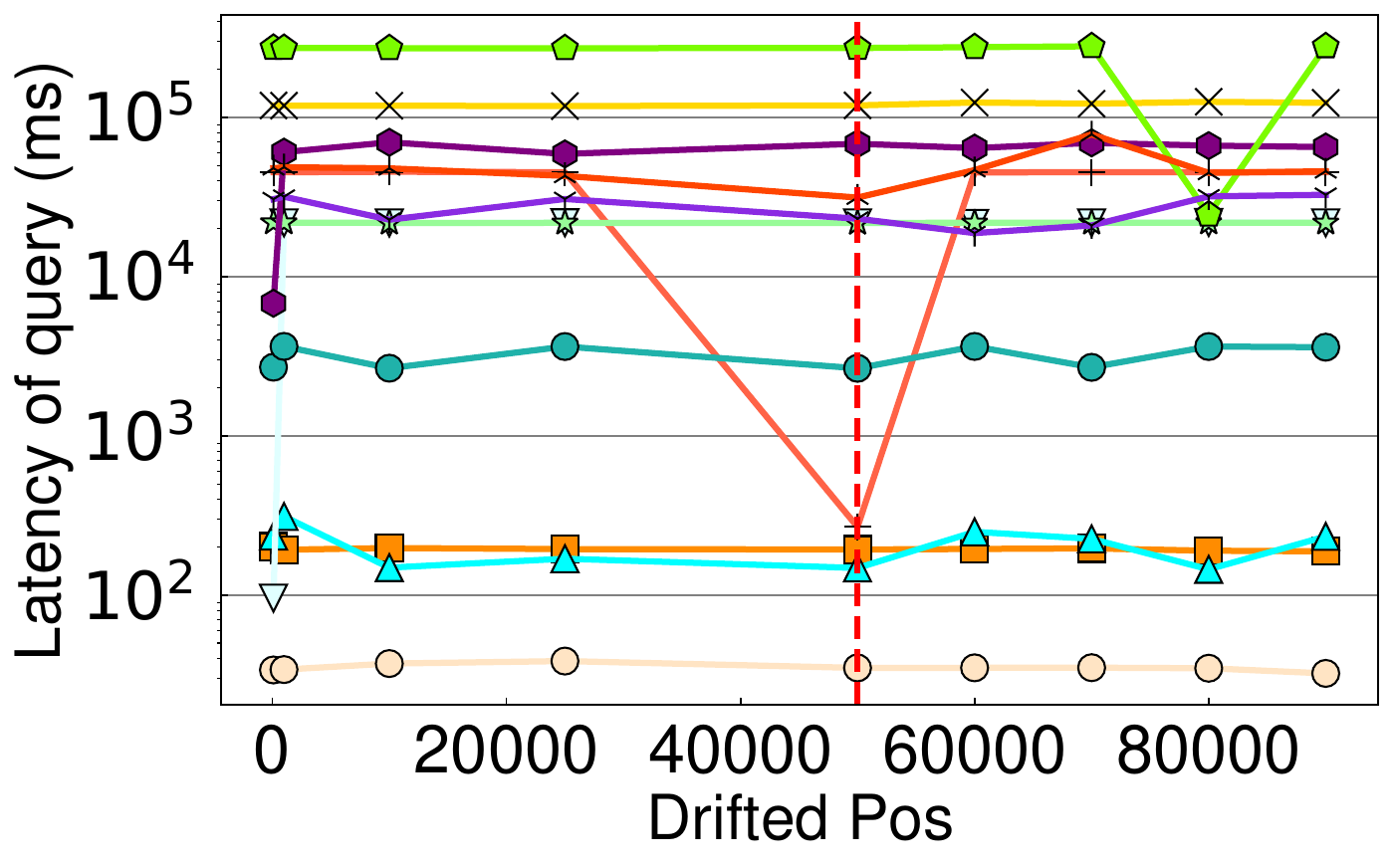}
            \label{fig:cd_lat}
        }
    \end{minipage}     
    \caption{Tuning the occurrence position of distribution drift. The Red dashed line indicates where the distribution drift occurs exactly at the same place as the beginning of online ingested data.}
        \label{fig:cd_pos}
    \end{minipage}
    \begin{minipage}[b]{0.29\textwidth}
        \includegraphics*[width=0.99\textwidth]{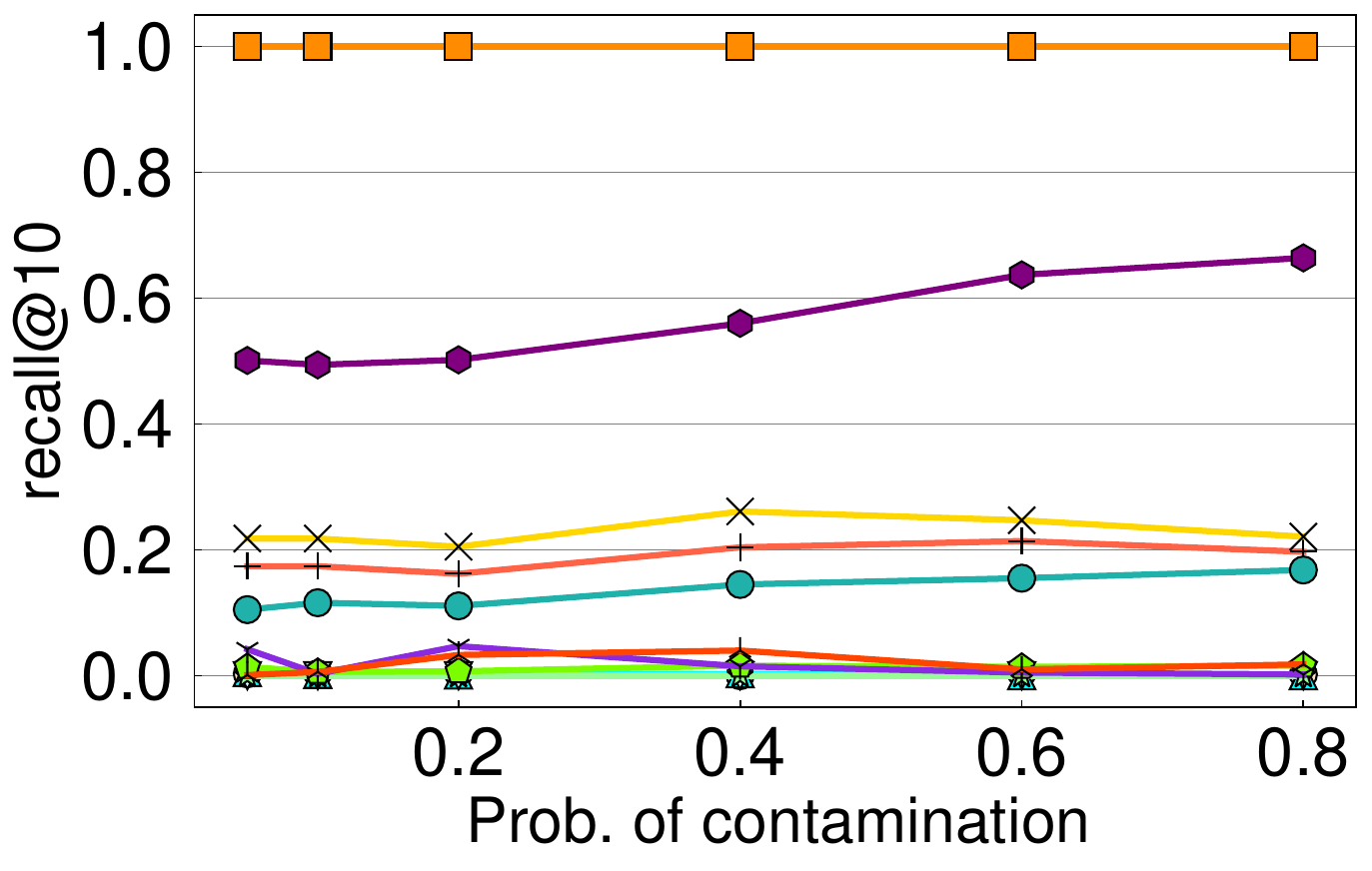}
        \caption{Tuning the intensity of distribution drift. A larger contamination probability indicates higher intensity.}
    \label{fig:cd_intensity}
    \end{minipage}    
\end{minipage}
\end{figure}

To investigate the impact of distribution drifts on AKNN algorithms, we varied the occurrence positions of drifted noun dictionary lists from 1,000 to 90,000. The resulting query recall and latency metrics are illustrated in Figure~\ref{fig:cd_pos}. Notably, drifts occurring before the 50,000 mark tend to be recognized during the offline initial data loading phase by AKNN algorithms, influencing their subsequent online performance.

Among the algorithms evaluated, \algoHNSW shows the highest resilience, maintaining a recall rate above 0.4, despite slight fluctuations in both latency and recall. These fluctuations arise primarily from how \algoHNSW adapts to the significant dissimilarities between embedding vectors before and after the drift. It may either create a new graph cluster for the drifted data, which is efficient and precise, or try to integrate it into the existing graph structure, which can lead to higher costs and inaccuracies due to unnecessary traversals through irrelevant graph vertices. This tension between integration and segregation is the main cause of the observed variability in its performance.

In contrast, all other evaluated AKNN algorithms struggle to achieve a recall rate above 0.1 when faced with distribution drifts. In particular, despite \algoOnlinePQ's design intentions to adapt dynamically to changing conditions, its simplistic update rule falls short, leading to suboptimal performance that is even less effective than \algoPQ. It fails to harmonize the original and new data distributions effectively, resulting in a compromised state where neither old nor new data is accurately processed or indexed. \algoIVFPQ, a ranging-based AKNN, shows substantial fluctuations in performance, especially when the drift occurs early in the data ingestion process. Its dependence on a fixed data distribution and pre-trained cluster centroids makes it particularly vulnerable to changes. This effect is exacerbated by its hierarchical design, which amplifies the fluctuation compared to \algoPQ.

We next investigated how varying the intensity of distribution drifts, by adjusting the probability of word contamination from 0 to 0.8, affects query recall. These results are presented in Figure~\ref{fig:cd_intensity}. The query latency, similar to evaluations on the \DPR dataset (Table~\ref{tab:e2e_ip}), showed marginal fluctuations across algorithms, with deviations no greater than 5\%. Across the board, most AKNN algorithms exhibited low recall rates and higher latency, with recalls consistently under 0.4, except for \algoHNSW. Notably, \algoHNSW's recall improved as the drift intensity increased, which can be credited to its robust hierarchical structure and adaptive heuristics. This algorithm effectively clusters drift-related data around hot-spot topics within its proxy graph, enhancing its capability to retrieve relevant entries. This process is optimized by its hierarchical design, which manages the scope of each cluster, reducing the likelihood of including irrelevant results and thus boosting accuracy. 

To further analyze \algoHNSW, we decompose its data ingestion process into three distinct parts: 1) \emph{Greedy}, which involves the greedy search to identify the insertion region; 2) \emph{Candidate}, which narrows down the connecting data points; and 3) \emph{Link}, which connects the selected candidates with the new data.
We examine different datasets, \DPR and \Wte, across various word contamination probabilities (ranging from 0 to 0.8), with the time breakdown illustrated in Figure~\ref{fig:hnsw_bd}. As expected, the distribution shift (\Wte) results in a larger proportion of time spent on \emph{Candidate} compared to the no shift scenario (\DPR), as identifying close data point regions becomes more challenging under the shift. Additionally, the proportion of time spent on \emph{Link} decreases when the data is less semantically relevant. Specifically, \DPR, generated from a real-world corpus, is more semantically relevant than \Wte, which comprises fabricated sentences. The semantic relevance of \Wte also increases slightly with higher word contamination.

%% file: EvaluationShort/Agglomeration.tex
\begin{figure}[t]
\begin{minipage}[a]{\textwidth} 
\centering
                \begin{minipage}[b]{0.29\textwidth}
        \includegraphics*[width=0.99\textwidth]{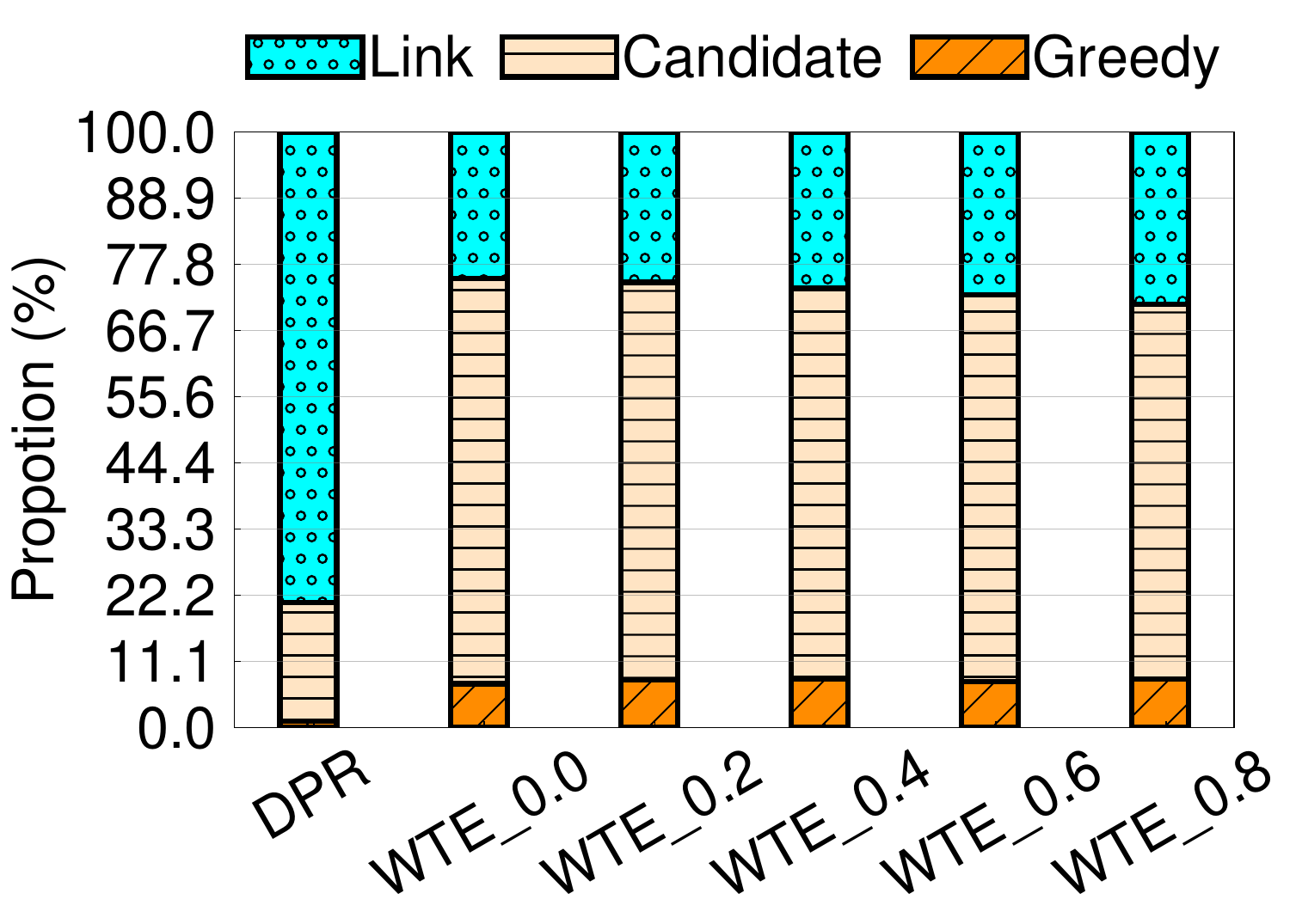}
        \caption{Breakdown of \algoHNSW on \DPR and \Wte.}
    \label{fig:hnsw_bd}
    \end{minipage}         \begin{minipage} [b] {0.70\textwidth}
        \begin{minipage}[c]{0.57\textwidth}
        \centering
        \subfigure[Query Recall]{
            \includegraphics[width=0.99\textwidth]{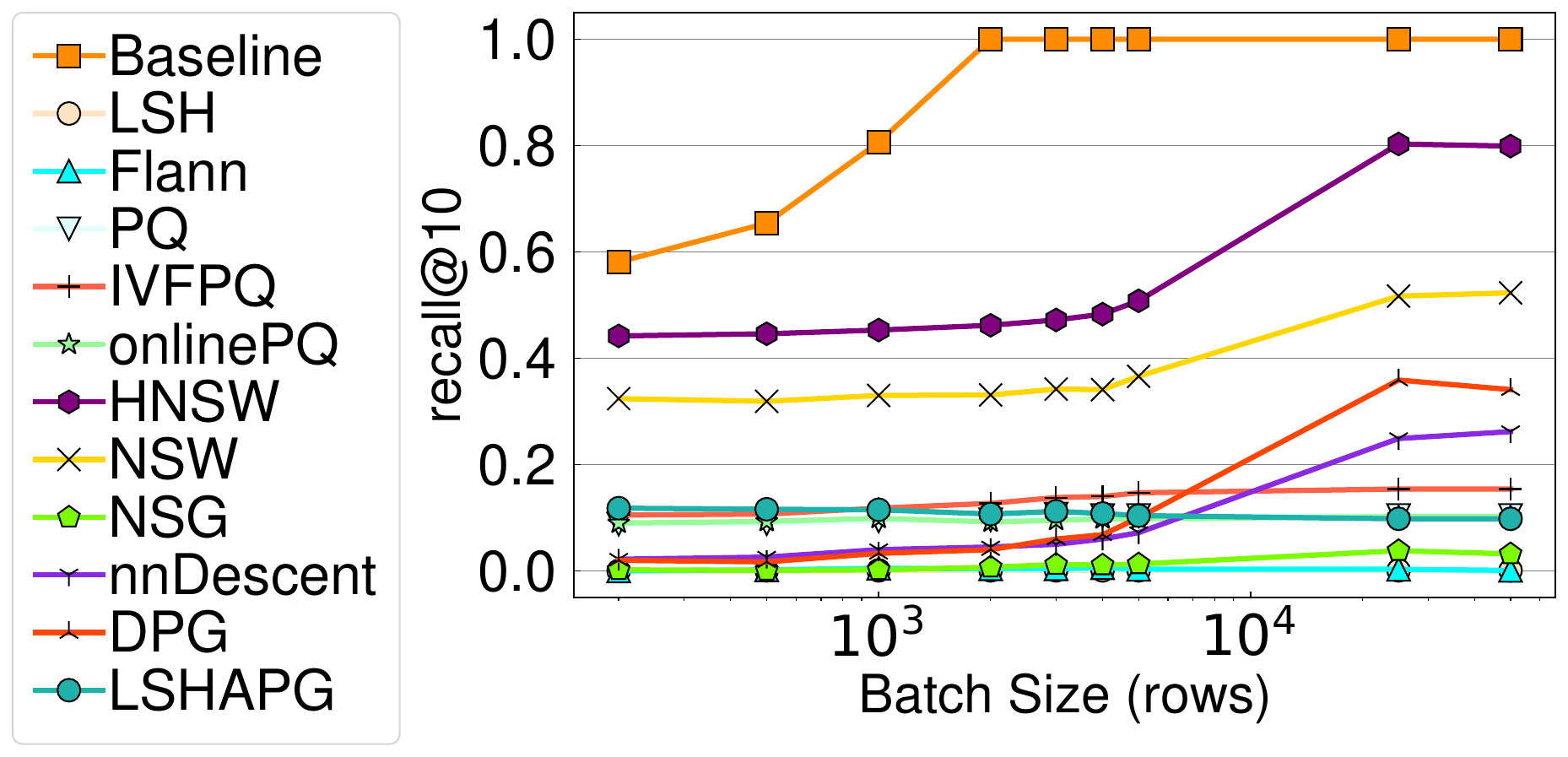}
            \label{fig:batch_recall}
        }
    \end{minipage}
    \begin{minipage}[c]{0.41\textwidth}
        \subfigure[Query Latency.]{
            \includegraphics[width=0.99\textwidth]{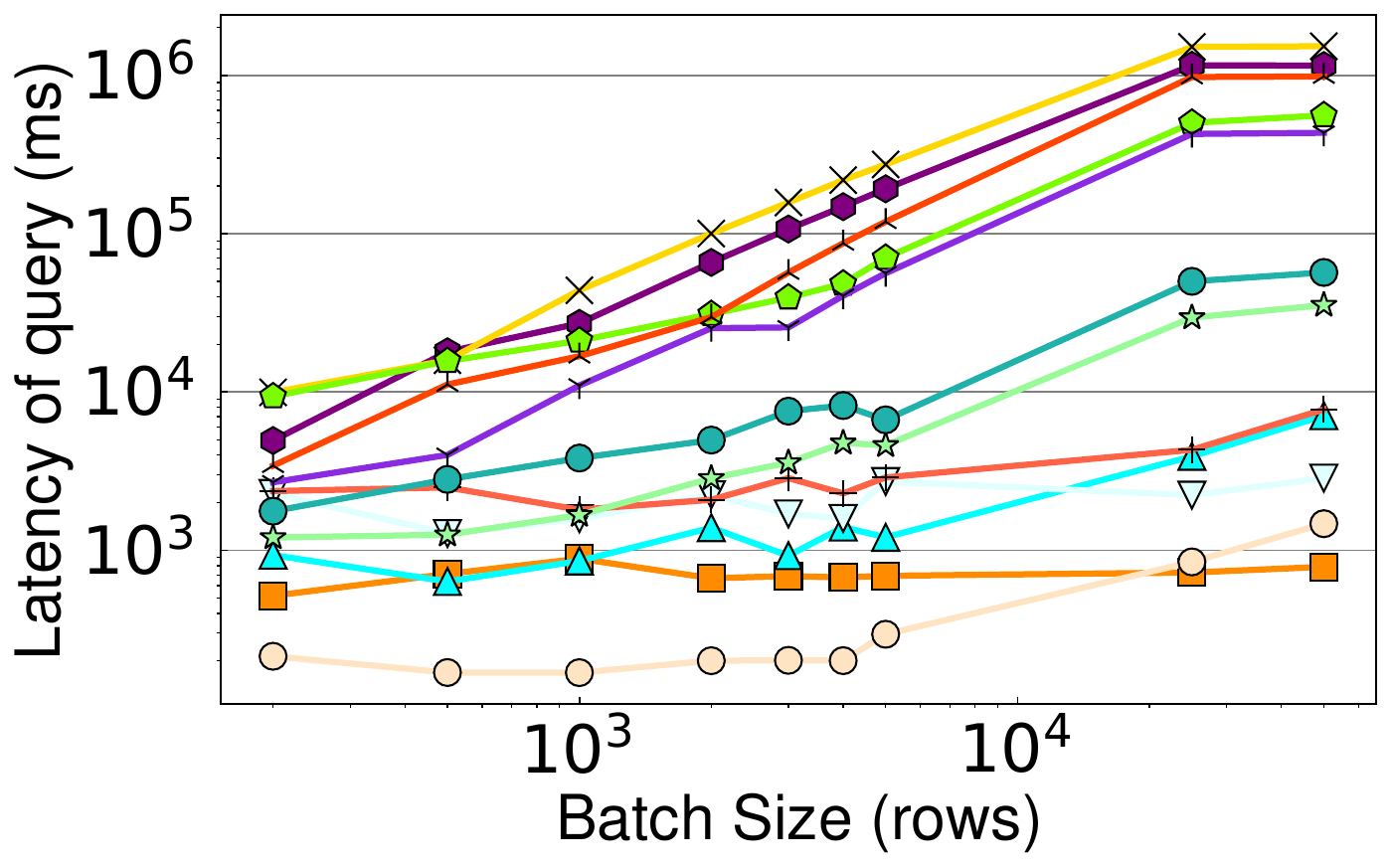}
            \label{fig:batch_lat}
        }
    \end{minipage}     
    \caption{Tuning the size of micro-batches from 200 to 50000. }
        \label{fig:batch}
    \end{minipage}
    
\end{minipage}
\end{figure}

\subsection{Micro-Batch Size Tuning is a Hard Game}
\label{subsec:batching_hard}
Our experiments with varying micro-batch sizes from 200 to 50,000 in the context of the \DPR datasets demonstrated notable impacts on query latency and recall, with results presented in Figure~\ref{fig:batch}. These findings underscore that there is no universally optimal batch size that suits all AKNN algorithms.

\begin{quote}
    Observation \themyvar.
    {Micro-batch size significantly affects AKNN performance, and the optimal setting varies by algorithm.}
\end{quote}
\stepcounter{myvar}
The efficiency of batch processing is highly dependent on the specific AKNN algorithm used. For algorithms that process vector points individually, such as \algoHNSW, \algoNSW, and \algoNSG, larger batch sizes tend to reduce processing efficiency, consequently increasing query latency as shown in Figure~\ref{fig:batch_lat}. In contrast, algorithms like \algoLSH demonstrate optimal performance at a moderate batch size—around 1,000—balancing the benefits of data aggregation with processor utilization.

Additionally, our experiments revealed that batch size profoundly impacts query recall, particularly through its influence on data ingestion efficiency and the likelihood of data droppings (Figure~\ref{fig:batch_recall}). Smaller batch sizes, while seemingly agile, often lead to higher data loss when system capacity is exceeded, as evidenced by increased inaccuracies in \algoBF’s performance, which is typically expected to execute exact searches. Larger batch sizes mitigate this risk by reducing the time spent segmenting ingested data and minimizing static overhead, thereby preserving more complete data sets and enhancing recall across all tested algorithms.

%% file: EvaluationShort/Opt.tex
\subsection{Optimizations are Promising, but There are Still Open Challenges.} 
\label{subsec:aknn_opt}
This section evaluates AKNN optimizations, encompassing both machine learning-based enhancements and optimized distance computation strategies. These optimizations are tested using representative algorithms \algoLSH and \algoHNSW, respectively, under conditions that include steady and shifted data distributions using datasets \DPR and \Wte. Please refer to Appendix~\ref{subsec:eva_setup} for more setup details. Results are summarized in Table~\ref{tab:combined_aknn}.

\begin{quote}
    Observation \themyvar.
    {Machine learning leads to large improvements of AKNN due to better data organization, but is challenged by distribution shifts.}
\end{quote}
\stepcounter{myvar}

\textbf{Machine learning} (ML) significantly improves AKNN performance through enhanced data organization, as observed with \algoLSH in our experiments. By integrating a simple MLP model for hashing, trained offline to mimic spectral hashing loss functions, we achieved a $7\times$ increase in accuracy. However, this method incurs higher query and vector search latencies due to its computational overhead. Interestingly, machine learning reduced pending writing latency by 66\%, optimizing vector assignments and thus improving overall efficiency. Despite these gains, machine learning struggles under dynamic conditions with data distribution shifts, leading to exhaustive searches across all buckets and even higher operational costs than \algoBF.

\begin{table}[]
\centering
\resizebox{0.7\textwidth}{!}{%
\begin{tabular}{|ll|llll|llll|}
\hline
\multicolumn{2}{|l|}{\multirow{2}{*}{Algorithms}}               & \multicolumn{4}{l|}{Not Drift}                                                                                          & \multicolumn{4}{l|}{With Drift}                                                                                         \\ \cline{3-10} 
\multicolumn{2}{|l|}{}                                          & \multicolumn{1}{l|}{Recall}         & \multicolumn{1}{l|}{QL}              & \multicolumn{1}{l|}{VSL}   & PWL           & \multicolumn{1}{l|}{Recall}        & \multicolumn{1}{l|}{QL}              & \multicolumn{1}{l|}{VSL}           & PWL    \\ \hline
\multicolumn{2}{|l|}{\algoBF}                                   & \multicolumn{1}{l|}{1.000}          & \multicolumn{1}{l|}{0.19}            & \multicolumn{1}{l|}{0.12}  & 0.07          & \multicolumn{1}{l|}{1.000}         & \multicolumn{1}{l|}{0.21}            & \multicolumn{1}{l|}{0.14}          & 0.07   \\ \hline
\multicolumn{1}{|l|}{\multirow{2}{*}{ML}}  & \algoLSH w/o ML    & \multicolumn{1}{l|}{\textbf{0.003}} & \multicolumn{1}{l|}{0.03}            & \multicolumn{1}{l|}{0.01}  & \textbf{0.03} & \multicolumn{1}{l|}{0.003}         & \multicolumn{1}{l|}{\textbf{0.04}}   & \multicolumn{1}{l|}{\textbf{0.01}} & 0.03   \\ \cline{2-10} 
\multicolumn{1}{|l|}{}                     & \algoLSH w/ ML     & \multicolumn{1}{l|}{\textbf{0.024}} & \multicolumn{1}{l|}{0.10}            & \multicolumn{1}{l|}{0.10}  & \textbf{0.01} & \multicolumn{1}{l|}{0.374}         & \multicolumn{1}{l|}{\textbf{5.82}}   & \multicolumn{1}{l|}{\textbf{5.82}} & 0.01   \\ \hline
\multicolumn{1}{|l|}{\multirow{3}{*}{DCO}} & \algoHNSW          & \multicolumn{1}{l|}{0.41}           & \multicolumn{1}{l|}{\textbf{485.53}} & \multicolumn{1}{l|}{47.78} & 437.75        & \multicolumn{1}{l|}{\textbf{0.48}} & \multicolumn{1}{l|}{113.82}          & \multicolumn{1}{l|}{11.26}         & 102.56 \\ \cline{2-10} 
\multicolumn{1}{|l|}{}                     & \algoHNSW w/ comp. & \multicolumn{1}{l|}{0.30}           & \multicolumn{1}{l|}{\textbf{99.32}}  & \multicolumn{1}{l|}{0.08}  & 99.24         & \multicolumn{1}{l|}{\textbf{0.09}} & \multicolumn{1}{l|}{15.41}           & \multicolumn{1}{l|}{0.07}          & 15.34  \\ \cline{2-10} 
\multicolumn{1}{|l|}{}                     & \algoHNSW w/ rand. & \multicolumn{1}{l|}{0.42}           & \multicolumn{1}{l|}{\textbf{87.58}}  & \multicolumn{1}{l|}{0.09}  & 87.49         & \multicolumn{1}{l|}{0.39}          & \multicolumn{1}{l|}{\textbf{123.51}} & \multicolumn{1}{l|}{11.45}         & 112.06 \\ \hline
\end{tabular}
}
\caption{Combined results for AKNN with machine learning (ML) and distance computation optimizations (DCO). ``QL'' is short for query latency, ``VSL'' is short for vector search latency, and ``PWL'' is short for pending writing latency. The unit of latency is $\times 1000ms$.}
\label{tab:combined_aknn}
\end{table}


\begin{quote}
    Observation \themyvar.
    {Optimized distance computation is also helpful, but more awareness should be put into solving the ingestion efficiency challenge.}
\end{quote}

\textbf{Distance computation optimizations} (DCO) include data compression (comp.) and randomized calculations (rand.). These methods proved beneficial, reducing query latency by up to 81.7\% when using compression techniques such as Locally-Adaptive Vector Quantization when applied to \algoHNSW in our study as an example. However, challenges persist with ingestion efficiency, preventing optimized AKNNs from surpassing the query latency performance of \algoBF. Inefficient data ingestion impacts the resilience of optimizations against data distribution shifts, as seen with compression where recall significantly dropped from 0.48 to 0.09 under shifted conditions.
Specifically, the countermeasures against distribution shifts like~\cite{aguerrebere2024locally} are hindered by an insufficient sample size due to data dropping, leading to their ineffectiveness. On the other hand, randomized calculation increases latency in comparison to \algoHNSW. This is because it tends to cause \algoHNSW to become mired in clusters of poorly ingested, distant data points when data dropping and distribution shifts coincide. This issue highlights the critical need for improving ingestion efficiency to leverage the full potential of AKNN optimizations in dynamic environments.

%% file: Limitations.tex
\section{Limitations}
\label{sec:limitation}
The limitations of \system are outlined as follows:
First, \system operates under single-threaded, in-memory, and CPU-only conditions, not yet harnessing advances such as SSDs \cite{singh2021freshdiskann} and GPUs \cite{li2023learning} that could boost AKNN performance. Exploring these technologies is a crucial direction for future research. 
Despite incorporating an implementation of parallel and distributed computing based on frameworks by Wang et al.~\cite{wang2021milvus} and Moritz et al.~\cite{moritz2018ray} in \system, no significant new insights have emerged due to the shared-nothing architecture of the data shards. 
Second, \system does not support dynamic deletion due to the complexities of data expiration and resource recycling as highlighted by Singh et al.~\cite{singh2021freshdiskann}. 
Efficiently managing and recycling outdated information within dynamic environments presents challenges that complicate the direct application of deletion processes, impacting both system performance and data integrity.
Third, the scope of \system's evaluation is 
not yet extended to the accuracy of downstream tasks like Retrieval-Augmented Generation (RAG). This limitation is due to the specialized nature of these downstream tasks, which involve additional complexities beyond the primary objectives of this benchmark. While these tasks are crucial, as noted in the works of Lyu et al. ~\cite{lyu2024crud} and Jiang et al. ~\cite{jiang2023chameleon}, addressing them requires a distinct set of methodologies and metrics that fall outside the current scope of \system.
Lastly, concerning data handling within \system, assumptions about data arrival are overly simplified—data is expected to arrive in order and at a constant speed. This setup does not account for more realistic, complex scenarios involving out-of-order or skewed data distributions that are increasingly common in real-world applications. Addressing these challenges, as explored by Heliou et al.~\cite{heliou2020gradient}, Zeng et al.~\cite{zeng2024pecj} and Zhang et al.~\cite{zhang2023scalable}, involves intricate strategies for error compensation and load balancing, which we aim to incorporate in future iterations of our benchmark.

%% file: Conclusion.tex
\section{Conclusion}
In this paper, we introduced \system, a benchmark that goes beyond traditional static data scenarios, offering a deep dive into the performance of AKNN algorithms under conditions of continuous data ingestion. \system is publicly available 
and is compatible with LibTorch/PyTorch, promoting collaborative innovation within the machine learning and data management communities.
Our experimental analysis using \system has shed light on unique challenges faced in dynamic data contexts, particularly the efficiency of data ingestion and the significant effects of data distribution shifts on AKNN performance. Furthermore, \system elucidates complex factors like micro-batching configurations and evaluates the extent and limitations of existing AKNN optimizations. 
The insights gained from this benchmark are poised to drive forward research into more efficient AKNN workflows and the development of advanced machine learning strategies better suited for managing data dynamics. We anticipate that the continuous evolution of \system will inspire ongoing improvements and adaptations in the field, addressing both current and emerging challenges in dynamic data handling.


%% file: Appendix.tex
\newpage
\input{Checklist}


\section{Appendix}

\subsection{Supplementary Details of \system Benchmark}
\label{subsec:detail_benchmark}
This section provides supplementary details of the \system, including a comprehensive overview of datasets, implementation specifics, and deployment considerations.

\begin{table}[h]
\centering
    \centering
    \resizebox{0.5\textwidth}{!}{%
   \begin{tabular}{|l|l|l|l|}
\hline
\multirow{7}{*}{Real-world} & \textbf{Name}                                                   & \textbf{Description} & \textbf{Dimension} \\ \cline{2-4} 
                            & \Glove~\cite{li2019approximate}                                 & Text embedding       & $100$              \\ \cline{2-4} 
                            & \SIFT~\cite{li2019approximate}                                  & Image                & $128$              \\ \cline{2-4} 
                            & \Msong~\cite{li2019approximate}                                 & Audio                & $4208$             \\ \cline{2-4} 
                            & \Sun~\cite{li2019approximate}                                   & Image                & $512$              \\ \cline{2-4} 
                            & \DPR~\cite{aguerrebere2024locally}                              & Text embedding       & $768$              \\ \cline{2-4} 
                            & \Trevi~\cite{li2019approximate}                                 & Image                & $4096$             \\ \hline
\multirow{2}{*}{Synthetic}  & \Random~\cite{Pytorch}                                          & i.i.d values         & Adjustable         \\ \cline{2-4} 
                            & \Wte~\cite{aguerrebere2024locally,warthunderWiki,loper2002nltk} & Text embedding       & $768$              \\ \hline
\end{tabular}
}
\caption{Datasets Summary.}
\label{tab:realworld_workloads}
\end{table}

\paragraph{More Details about Datasets}
Table~\ref{tab:realworld_workloads} summarizes the datasets. The \DPR dataset was reconstructed using the source code and corpus provided by Aguerrebere et al.~\cite{aguerrebere2024locally}, while the \Glove, \SIFT, \Msong, \Sun, and \Trevi datasets were obtained from Li et al.~\cite{li2019approximate}. Each real-world dataset is normalized using the \emph{torch::norm} function from LibTorch~\cite{Pytorch} to ensure values range between -1 and 1. The \Random dataset is generated by the \emph{torch::random} function, producing a sequence of independent and identically distributed (i.i.d.) uniform values between $0$ and $1$. This dataset allows for configurable dimensions and volumes of vectors, facilitating studies on dimensional scalability. In our experiments, we use \DPR if not otherwise specified. By default, 
we initially loaded AKNN with 50K rows of vector in the offline stage. Afterwards, we enter the dynamic ingestion phase by feeding AKNN with 50K new data (by default 4K rows/second), and let the micro-batches sized $4K$.  

The \Wte dataset simulates text embedding processes similar to \DPR but uses synthetically generated text sentences. Each sentence comprises a subject, predicate, and object, with both subject and object as nouns, and predicates as random verbs from NLTK~\cite{loper2002nltk}, ensuring no biased distribution. Controlled data distribution shifts are introduced by varying the noun categories, for instance, comparing nouns from USSR land vehicles against USA aircrafts from Warthunder~\cite{warthunderWiki}. The embedding models used are static, without fine-tuning, set to produce 768-dimensional vectors for 100K rows. Additionally, we manipulate the occurrence position and intensity of distribution shifts in \Wte. For occurrence position, we adjust where vector embeddings transition from USA aircraft nouns to USSR land vehicle nouns, with word contamination set to a probability of 1 thereafter. To vary the shift intensity, we construct sentences using USA aircraft nouns, randomly replace the subject noun with ``T-34-85,'' and generate embeddings from these contaminated sentences. This models real-world scenarios of emerging topics, with the ``T-34-85'' representing a sudden shift in focus. The contamination probability is adjustable to simulate varying intensities of this shift.


\paragraph{More Implementation Details} 
We have standardized the API access for all AKNN algorithms within \system to include initialization processes, offline data loading, data ingestion, and vector search, ensuring full compatibility with LibTorch and PyTorch~\cite{Pytorch}, popular tools in the data science community, across both C++ and Python. This unified approach extends through our C++ codebase, utilizing static compilation and interfaces based on \emph{torch::Tensor} to maintain consistency and compatibility across experiments. Our implementation adheres to the IEEE 754 32-bit floating-point (FP32) format for vector elements and leverages \emph{torch::Tensor} for vector representation, thereby aligning with LibTorch and PyTorch without requiring extra conversions. All code and scripts are publicly available at \url{https://github.com/intellistream/candy}, with additional documentation provided in our codebase. 

\paragraph{Parallel Hardware Considerations}
Our experiments primarily utilized a Silver 4310 processor. We adapted approximate k-nearest neighbor (AKNN) algorithms for parallel and distributed computing, employing a straightforward sharding approach as outlined in Wang et al.\cite{wang2021milvus} and Moritz et al.\cite{moritz2018ray}. However, this embarrassingly parallel approach did not yield novel insights due to its inherent shared-nothing architecture. We also refrained from implementing algorithm-specific parallelization techniques, such as concurrent vertex writing in \algoHNSW~\cite{singh2021freshdiskann,aguerrebere2024locally}, owing to the complexity of ensuring fairness in such configurations. Therefore, our evaluation emphasizes single-threaded performance to ensure consistency across tests.

\subsection{More Related Work Discussion}
\label{subsec:detail_algorithm}
\paragraph{More Discussions on Ranging-based AKNN}
Locality Sensitive Hashing (\algoLSH~\cite{gionis1999similarity}), one of the earliest ranging-based AKNNs, hashes similar vectors into the same bucket to form ranges, though it often faces criticism for its suboptimal trade-off between retrieval efficiency and accuracy. Subsequent ranging-based AKNN methods have sought to improve this by employing more sophisticated techniques like product quantization (\algoPQ~\cite{jegou2010product}) and space partitioning (\algoFlann~\cite{muja2014scalable}), which consider the distribution of data collection \vectorSet. Enhancements such as hierarchical range layouts in \algoIVFPQ~\cite{jegou2010product}, which uses two tiers of quantizers to refine \algoPQ, further boost retrieval efficiency. Despite their streamlined workflow for data retrieval and ingestion, ranging-based AKNNs often struggle with accurately determining ranges, which can misplace distant vectors in the same range or similar vectors in separate ranges. Most rely on the assumption that \vectorSet's distribution is known and static, except for \algoOnlinePQ, which incrementally updates its cluster centroids to adapt to distribution shifts in newly ingested data~\cite{xu2018online}.

\paragraph{More Discussions on Navigation-based AKNN}
Navigation-based AKNN employs various strategies to manage the navigation path and decide when to halt the expansion of candidate neighbors. For instance, \algoNSW~\cite{malkov2014approximate} uses the Delaunay Graph strategy, while \algoNNDecent~\cite{dong2011efficient} relies on the K-Nearest Neighbor Graph. Both \algoDPG~\cite{li2019approximate} and \algoNSG~\cite{fu2019fast} utilize the K-Nearest Neighbor Graph and Relative Neighborhood Graph approaches, with \algoNSG enhancing \algoDPG’s retrieval efficiency through different vertex assignment techniques. Additionally, hierarchical designs similar to those in ranging-based AKNN are also used in navigation-based methods. For example, \algoHNSW~\cite{malkov2018efficient} builds multiple layers of \algoNSW, significantly improving retrieval efficiency by avoiding the flat structure limitations. 
Furthermore, the graph maintenance can also be optimized by using hashing, as recently proposed by \algoLSHAPG~\cite{zhao2023towards}.
Navigation-based AKNN effectively overcomes the challenges of imprecise range determination found in ranging-based methods, achieving higher retrieval accuracy and efficiency. However, modifying navigation paths within a proxy graph is a complex, iterative process, often requiring adjustments among neighboring vectors.

%% file: Checklist.tex
\section*{Checklist}


\begin{enumerate}

\item For all authors...
\begin{enumerate}
  \item Do the main claims made in the abstract and introduction accurately reflect the paper's contributions and scope?
    \answerYes{}
  \item Did you describe the limitations of your work?
    \answerYes{In Section~\ref{sec:limitation}.}
  \item Did you discuss any potential negative societal impacts of your work?
    \answerNo{It's hard for us to consider the negative social impacts of AKNN, which is not an application but a fundamental data science operator.}
  \item Have you read the ethics review guidelines and ensured that your paper conforms to them?
    \answerYes{}
\end{enumerate}

\item If you are including theoretical results...
\begin{enumerate}
  \item Did you state the full set of assumptions of all theoretical results?
    \answerNA{}
	\item Did you include complete proofs of all theoretical results?
    \answerNA{}
\end{enumerate}

\item If you ran experiments (e.g. for benchmarks)...
\begin{enumerate}
  \item Did you include the code, data, and instructions needed to reproduce the main experimental results (either in the supplemental material or as a URL)?
    \answerYes{At \url{https://github.com/intellistream/candy}.}
  \item Did you specify all the training details (e.g., data splits, hyperparameters, how they were chosen)?
    \answerNA{Model training is not within our major scope, but we mentioned the details in our Appendix~\ref{subsec:eva_setup} when machine learning is involved.}
	\item Did you report error bars (e.g., with respect to the random seed after running experiments multiple times)?
    \answerNo{The result in our running is stable multiple times.}
	\item Did you include the total amount of compute and the type of resources used (e.g., type of GPUs, internal cluster, or cloud provider)?
    \answerYes{Single-thread, CPU only (Section~\ref{sec:limitation}).}
\end{enumerate}

\item If you are using existing assets (e.g., code, data, models) or curating/releasing new assets...
\begin{enumerate}
  \item If your work uses existing assets, did you cite the creators?
    \answerYes{}
  \item Did you mention the license of the assets?
    \answerNo{Some of them like \Glove are just shipped under Dropbox links from creators, we included the license of the rest in our code base (i.e., \DPR and \Random).}
  \item Did you include any new assets either in the supplemental material or as a URL?
    \answerYes{\Wte dataset in Appendix~\ref{subsec:detail_benchmark}.}
  \item Did you discuss whether and how consent was obtained from people whose data you're using/curating?
    \answerNA{The data is a free Wiki and public resource.}
  \item Did you discuss whether the data you are using/curating contains personally identifiable information or offensive content?
    \answerNo{While there is no personally identifiable information included, it is challenging to ensure that every term, such as aircraft nicknames, used in generating embeddings, is not potentially offensive to someone.}
\end{enumerate}

\item If you used crowdsourcing or conducted research with human subjects...
\begin{enumerate}
  \item Did you include the full text of instructions given to participants and screenshots, if applicable?
    \answerNA{}
  \item Did you describe any potential participant risks, with links to Institutional Review Board (IRB) approvals, if applicable?
    \answerNA{}
  \item Did you include the estimated hourly wage paid to participants and the total amount spent on participant compensation?
    \answerNA{}
\end{enumerate}

\end{enumerate}
\newpage

%% file: Evaluations/Extra.tex
\subsection{Supplementary Experimental Insights}
\label{subsec:eva_sup}
This section presents additional insights gained from using \system. Specifically, we highlight significant issues such as the queuing of uningested data and re-examine the well-documented ``curse of dimension''~\cite{li2019approximate} in the context of dynamic data ingestion. Additional findings concerning data sources and data volumes are also discussed.

\begin{table}[]
 \resizebox{0.99\textwidth}{!}{%
\begin{tabular}{|ll|llllll|llllll|}
\hline
\multicolumn{2}{|l|}{\multirow{2}{*}{Algorithms}}                                & \multicolumn{6}{l|}{Recall@10}                                                                                                                                                                         & \multicolumn{6}{l|}{Query Latency ($\times 1000 ms$)}                                                                                                                                                                \\ \cline{3-14} 
\multicolumn{2}{|l|}{}                                                           & \multicolumn{1}{l|}{\Glove}        & \multicolumn{1}{l|}{\SIFT}         & \multicolumn{1}{l|}{\Msong}        & \multicolumn{1}{l|}{\Sun}          & \multicolumn{1}{l|}{\DPR}          & \Trevi        & \multicolumn{1}{l|}{\Glove}          & \multicolumn{1}{l|}{\SIFT}           & \multicolumn{1}{l|}{\Msong}          & \multicolumn{1}{l|}{\Sun}             & \multicolumn{1}{l|}{\DPR}             & \Trevi          \\ \hline
\multicolumn{2}{|l|}{\textbf{\algoBF}}                                           & \multicolumn{1}{l|}{\textbf{1.00}} & \multicolumn{1}{l|}{\textbf{1.00}} & \multicolumn{1}{l|}{\textbf{1.00}} & \multicolumn{1}{l|}{\textbf{1.00}} & \multicolumn{1}{l|}{\textbf{1.00}} & \textbf{1.00} & \multicolumn{1}{l|}{\textbf{0.43}}   & \multicolumn{1}{l|}{\textbf{0.44}}   & \multicolumn{1}{l|}{\textbf{0.59}}   & \multicolumn{1}{l|}{\textbf{0.70}}    & \multicolumn{1}{l|}{\textbf{0.74}}    & \textbf{2.54}   \\ \hline
\multicolumn{1}{|l|}{\multirow{5}{*}{Ranging-based}}    & \textbf{\algoLSH}      & \multicolumn{1}{l|}{\textbf{0.00}} & \multicolumn{1}{l|}{\textbf{0.01}} & \multicolumn{1}{l|}{\textbf{0.00}} & \multicolumn{1}{l|}{\textbf{0.00}} & \multicolumn{1}{l|}{\textbf{0.00}} & \textbf{0.07} & \multicolumn{1}{l|}{\textbf{0.21}}   & \multicolumn{1}{l|}{\textbf{0.23}}   & \multicolumn{1}{l|}{\textbf{0.23}}   & \multicolumn{1}{l|}{\textbf{0.42}}    & \multicolumn{1}{l|}{\textbf{0.20}}    & \textbf{12.33}  \\ \cline{2-14} 
\multicolumn{1}{|l|}{}                                  & \algoFlann             & \multicolumn{1}{l|}{0.01}          & \multicolumn{1}{l|}{0.01}          & \multicolumn{1}{l|}{0.00}          & \multicolumn{1}{l|}{0.09}          & \multicolumn{1}{l|}{0.02}          & 0.13          & \multicolumn{1}{l|}{0.02}            & \multicolumn{1}{l|}{0.02}            & \multicolumn{1}{l|}{0.02}            & \multicolumn{1}{l|}{0.02}             & \multicolumn{1}{l|}{0.05}             & 0.08            \\ \cline{2-14} 
\multicolumn{1}{|l|}{}                                  & \algoPQ                & \multicolumn{1}{l|}{0.12}          & \multicolumn{1}{l|}{0.30}          & \multicolumn{1}{l|}{0.75}          & \multicolumn{1}{l|}{0.53}          & \multicolumn{1}{l|}{0.10}          & 0.53          & \multicolumn{1}{l|}{2.44}            & \multicolumn{1}{l|}{3.68}            & \multicolumn{1}{l|}{8.83}            & \multicolumn{1}{l|}{373.37}           & \multicolumn{1}{l|}{5.63}             & 22.14           \\ \cline{2-14} 
\multicolumn{1}{|l|}{}                                  & \algoIVFPQ             & \multicolumn{1}{l|}{0.04}          & \multicolumn{1}{l|}{0.23}          & \multicolumn{1}{l|}{0.02}          & \multicolumn{1}{l|}{0.33}          & \multicolumn{1}{l|}{0.15}          & 0.42          & \multicolumn{1}{l|}{4.11}            & \multicolumn{1}{l|}{6.19}            & \multicolumn{1}{l|}{18.94}           & \multicolumn{1}{l|}{5.66}             & \multicolumn{1}{l|}{16.97}            & 22.22           \\ \cline{2-14} 
\multicolumn{1}{|l|}{}                                  & \textbf{\algoOnlinePQ} & \multicolumn{1}{l|}{\textbf{0.12}} & \multicolumn{1}{l|}{\textbf{0.31}} & \multicolumn{1}{l|}{\textbf{0.82}} & \multicolumn{1}{l|}{\textbf{0.52}} & \multicolumn{1}{l|}{\textbf{0.10}} & \textbf{0.52} & \multicolumn{1}{l|}{\textbf{2.53}}   & \multicolumn{1}{l|}{\textbf{1.21}}   & \multicolumn{1}{l|}{\textbf{16.86}}  & \multicolumn{1}{l|}{\textbf{399.04}}  & \multicolumn{1}{l|}{\textbf{29.81}}   & \textbf{162.16} \\ \hline
\multicolumn{1}{|l|}{\multirow{6}{*}{Navigation-based}} & \textbf{\algoHNSW}     & \multicolumn{1}{l|}{\textbf{0.60}} & \multicolumn{1}{l|}{\textbf{0.90}} & \multicolumn{1}{l|}{\textbf{1.00}} & \multicolumn{1}{l|}{\textbf{0.99}} & \multicolumn{1}{l|}{\textbf{0.80}} & \textbf{N.A.} & \multicolumn{1}{l|}{\textbf{733.55}} & \multicolumn{1}{l|}{\textbf{178.88}} & \multicolumn{1}{l|}{\textbf{560.32}} & \multicolumn{1}{l|}{\textbf{4411.18}} & \multicolumn{1}{l|}{\textbf{1008.13}} & \textbf{N.A.}   \\ \cline{2-14} 
\multicolumn{1}{|l|}{}                                  & \algoNSW               & \multicolumn{1}{l|}{0.14}          & \multicolumn{1}{l|}{0.55}          & \multicolumn{1}{l|}{0.34}          & \multicolumn{1}{l|}{0.78}          & \multicolumn{1}{l|}{0.52}          & N.A.          & \multicolumn{1}{l|}{1722.13}         & \multicolumn{1}{l|}{230.68}          & \multicolumn{1}{l|}{2659.95}         & \multicolumn{1}{l|}{1389.90}          & \multicolumn{1}{l|}{1349.35}          & N.A.            \\ \cline{2-14} 
\multicolumn{1}{|l|}{}                                  & \algoNSG               & \multicolumn{1}{l|}{N.A.}          & \multicolumn{1}{l|}{0.01}          & \multicolumn{1}{l|}{0.00}          & \multicolumn{1}{l|}{0.04}          & \multicolumn{1}{l|}{0.01}          & 0.03          & \multicolumn{1}{l|}{N.A.}            & \multicolumn{1}{l|}{0.02}            & \multicolumn{1}{l|}{0.03}            & \multicolumn{1}{l|}{0.02}             & \multicolumn{1}{l|}{0.06}             & 0.11            \\ \cline{2-14} 
\multicolumn{1}{|l|}{}                                  & \algoNNDecent          & \multicolumn{1}{l|}{0.01}          & \multicolumn{1}{l|}{0.12}          & \multicolumn{1}{l|}{0.00}          & \multicolumn{1}{l|}{0.23}          & \multicolumn{1}{l|}{0.17}          & 0.46          & \multicolumn{1}{l|}{267.26}          & \multicolumn{1}{l|}{591.39}          & \multicolumn{1}{l|}{390.37}          & \multicolumn{1}{l|}{340.39}           & \multicolumn{1}{l|}{556.66}           & 734.74          \\ \cline{2-14} 
\multicolumn{1}{|l|}{}                                  & \algoDPG               & \multicolumn{1}{l|}{0.08}          & \multicolumn{1}{l|}{0.33}          & \multicolumn{1}{l|}{0.06}          & \multicolumn{1}{l|}{0.30}          & \multicolumn{1}{l|}{0.31}          & 0.48          & \multicolumn{1}{l|}{7639.13}         & \multicolumn{1}{l|}{1134.88}         & \multicolumn{1}{l|}{5009.27}         & \multicolumn{1}{l|}{906.83}           & \multicolumn{1}{l|}{1335.47}          & 4308.02         \\ \cline{2-14} 
\multicolumn{1}{|l|}{}                                  & \algoLSHAPG            & \multicolumn{1}{l|}{0.07}          & \multicolumn{1}{l|}{0.10}          & \multicolumn{1}{l|}{0.59}          & \multicolumn{1}{l|}{0.24}          & \multicolumn{1}{l|}{0.10}          & 0.24          & \multicolumn{1}{l|}{31.16}           & \multicolumn{1}{l|}{24.67}           & \multicolumn{1}{l|}{41.81}           & \multicolumn{1}{l|}{53.84}            & \multicolumn{1}{l|}{52.31}            & 180.05          \\ \hline
\end{tabular}
}
\caption{Comparing AKNN algorithms with dynamic data ingestion when queuing uningested data. N.A. means that the evaluation of certain algorithms and certain datasets exceeds 4h wall time, and therefore unable to get further results.}
\label{tab:e2e_que}
\end{table}
\begin{table}[]
 \resizebox{0.99\textwidth}{!}{%
\begin{tabular}{|ll|lll|lll|lll|}
\hline
\multicolumn{2}{|l|}{\multirow{2}{*}{Algorithms}}                                & \multicolumn{3}{l|}{Propotion of PWL (\%)}                                                 & \multicolumn{3}{l|}{PWL ($\times 1000 ms$)}                                                    & \multicolumn{3}{l|}{VSL ($\times 1000 ms$)}                                              \\ \cline{3-11} 
\multicolumn{2}{|l|}{}                                                           & \multicolumn{1}{l|}{\Msong}         & \multicolumn{1}{l|}{\DPR}           & \Trevi         & \multicolumn{1}{l|}{\Msong}          & \multicolumn{1}{l|}{\DPR}             & \Trevi          & \multicolumn{1}{l|}{\Msong}        & \multicolumn{1}{l|}{\DPR}          & \Trevi         \\ \hline
\multicolumn{2}{|l|}{\textbf{\algoBF}}                                           & \multicolumn{1}{l|}{\textbf{73.04}} & \multicolumn{1}{l|}{\textbf{0.00}}  & \textbf{60.86} & \multicolumn{1}{l|}{\textbf{0.44}}   & \multicolumn{1}{l|}{\textbf{0.50}}    & \textbf{1.50}   & \multicolumn{1}{l|}{\textbf{0.15}} & \multicolumn{1}{l|}{\textbf{0.24}} & \textbf{1.03}  \\ \hline
\multicolumn{1}{|l|}{\multirow{5}{*}{Ranging-based}}    & \textbf{\algoLSH}      & \multicolumn{1}{l|}{\textbf{74.71}} & \multicolumn{1}{l|}{\textbf{88.03}} & \textbf{1.12}  & \multicolumn{1}{l|}{\textbf{0.17}}   & \multicolumn{1}{l|}{\textbf{0.18}}    & \textbf{0.13}   & \multicolumn{1}{l|}{\textbf{0.06}} & \multicolumn{1}{l|}{\textbf{0.02}} & \textbf{12.20} \\ \cline{2-11} 
\multicolumn{1}{|l|}{}                                  & \algoFlann             & \multicolumn{1}{l|}{98.90}          & \multicolumn{1}{l|}{99.86}          & 99.24          & \multicolumn{1}{l|}{1.69}            & \multicolumn{1}{l|}{37.89}            & 9.95            & \multicolumn{1}{l|}{1.71}          & \multicolumn{1}{l|}{37.94}         & 10.02          \\ \cline{2-11} 
\multicolumn{1}{|l|}{}                                  & \algoPQ                & \multicolumn{1}{l|}{97.12}          & \multicolumn{1}{l|}{95.36}          & 98.13          & \multicolumn{1}{l|}{8.58}            & \multicolumn{1}{l|}{5.38}             & 21.87           & \multicolumn{1}{l|}{0.25}          & \multicolumn{1}{l|}{0.25}          & 0.26           \\ \cline{2-11} 
\multicolumn{1}{|l|}{}                                  & \algoIVFPQ             & \multicolumn{1}{l|}{97.75}          & \multicolumn{1}{l|}{96.05}          & 91.11          & \multicolumn{1}{l|}{18.68}           & \multicolumn{1}{l|}{16.64}            & 20.24           & \multicolumn{1}{l|}{0.26}          & \multicolumn{1}{l|}{0.33}          & 1.99           \\ \cline{2-11} 
\multicolumn{1}{|l|}{}                                  & \textbf{\algoOnlinePQ} & \multicolumn{1}{l|}{\textbf{98.47}} & \multicolumn{1}{l|}{\textbf{99.16}} & \textbf{99.85} & \multicolumn{1}{l|}{\textbf{16.61}}  & \multicolumn{1}{l|}{\textbf{29.57}}   & \textbf{161.91} & \multicolumn{1}{l|}{\textbf{0.25}} & \multicolumn{1}{l|}{\textbf{0.25}} & \textbf{0.25}  \\ \hline
\multicolumn{1}{|l|}{\multirow{6}{*}{Navigation-based}} & \textbf{\algoHNSW}     & \multicolumn{1}{l|}{\textbf{99.99}} & \multicolumn{1}{l|}{\textbf{99.99}} & \textbf{N.A.}  & \multicolumn{1}{l|}{\textbf{560.26}} & \multicolumn{1}{l|}{\textbf{1008.00}} & \textbf{N.A.}   & \multicolumn{1}{l|}{\textbf{0.05}} & \multicolumn{1}{l|}{\textbf{0.13}} & \textbf{N.A.}  \\ \cline{2-11} 
\multicolumn{1}{|l|}{}                                  & \algoNSW               & \multicolumn{1}{l|}{100.00}         & \multicolumn{1}{l|}{99.98}          & N.A.           & \multicolumn{1}{l|}{2659.85}         & \multicolumn{1}{l|}{1349.03}          & N.A.            & \multicolumn{1}{l|}{0.09}          & \multicolumn{1}{l|}{0.32}          & N.A.           \\ \cline{2-11} 
\multicolumn{1}{|l|}{}                                  & \algoNSG               & \multicolumn{1}{l|}{99.99}          & \multicolumn{1}{l|}{99.99}          & 99.99          & \multicolumn{1}{l|}{388.36}          & \multicolumn{1}{l|}{465.32}           & 929.59          & \multicolumn{1}{l|}{388.39}        & \multicolumn{1}{l|}{465.39}        & 929.71         \\ \cline{2-11} 
\multicolumn{1}{|l|}{}                                  & \algoNNDecent          & \multicolumn{1}{l|}{99.81}          & \multicolumn{1}{l|}{99.86}          & 99.85          & \multicolumn{1}{l|}{389.62}          & \multicolumn{1}{l|}{555.88}           & 733.62          & \multicolumn{1}{l|}{0.76}          & \multicolumn{1}{l|}{0.78}          & 1.13           \\ \cline{2-11} 
\multicolumn{1}{|l|}{}                                  & \algoDPG               & \multicolumn{1}{l|}{99.96}          & \multicolumn{1}{l|}{99.91}          & 99.92          & \multicolumn{1}{l|}{5007.04}         & \multicolumn{1}{l|}{1334.21}          & 4304.74         & \multicolumn{1}{l|}{2.23}          & \multicolumn{1}{l|}{1.26}          & 3.27           \\ \cline{2-11} 
\multicolumn{1}{|l|}{}                                  & \algoLSHAPG            & \multicolumn{1}{l|}{99.39}          & \multicolumn{1}{l|}{98.94}          & 99.43          & \multicolumn{1}{l|}{41.55}           & \multicolumn{1}{l|}{51.75}            & 179.02          & \multicolumn{1}{l|}{0.25}          & \multicolumn{1}{l|}{0.56}          & 1.03           \\ \hline
\end{tabular}
}
\caption{Break down the query latency into pending write latency and vector search latency when queuing uningested data. ``PWL'' is short for pending writing latency, and ``VSL'' is short for vector search latency. N.A. means that the evaluation of certain algorithms and certain datasets exceeds 4h wall time, and therefore unable to get further results.}
\label{tab:e2e_bd_cross}
\end{table}

\paragraph{It's not Practically Useful to Queue Uningested Data Due to Large Extra Latency}
Contrary to the default drop-on-congestion approach, we experimented with storing congested data in an in-memory queue, allowing AKNN to process all incoming data eventually. However,  our tests show that queuing decreases ingestion efficiency, contradicting its assumed practicality. In cross-validation tests reported in Table~\ref{tab:e2e_que} and~\ref{tab:e2e_bd_cross}, we maintained all settings identical to those in Tables~\ref{tab:e2e_ip} and \ref{tab:e2e_bd}, except that the data source was paused to ensure complete data ingestion. Notably, while avoiding data dropping does increase recall—for example, \algoHNSW’s recall improved from 0.61 to 1.00 in processing the \Msong dataset—this method also significantly increases vector search latency and overall query processing time. For instance, the vector search latency for \algoOnlinePQ in the \DPR dataset decreased dramatically from 3.43s to 0.25s due to fewer traversals through distant vectors when the complete dataset is ingested. However, the total latency for queries, such as those handled by \algoHNSW for the \DPR dataset, increased from approximately 691s to 1008s. This increase is due to intrinsic delays in updating AKNN compared to the data source, with added costs from memory operations and cache misses, which significantly enlarge end-to-end query latency.

\paragraph{The ``Curse of Dimension'' is Compounded in Dynamic Ingestion}
Scalability concerning vector dimensions remains a pivotal issue for AKNN, as underscored in previous studies~\cite{li2019approximate}. In our exploration with the \Random dataset, where dimensions ranged from 64 to 4096, we observed notable insights. Consistent with static data evaluations~\cite{li2019approximate}, navigation-based AKNNs such as \algoHNSW generally surpass ranging-based AKNNs like \algoIVFPQ in recall, due to their efficient preservation of data relationships and minimal assumptions, albeit at a cost of increased memory usage leading to out-of-memory issues. Contrary to established expectations~\cite{li2019approximate,malkov2018efficient}, apart from \algoLSH, all tested AKNNs displayed a greater susceptibility to the ``curse of dimension,'' with pronounced increases in query latency as dimensions expanded. This contradicts the belief that AKNNs inherently minimize performance degradation with increased dimensions~\cite{jegou2010product,malkov2018efficient}, revealing an over-optimization in vector search capabilities at the expense of ingestion efficiency, and highlights a critical area for improvement.

\begin{figure}[t]
\begin{minipage}[a]{\textwidth} 
\centering
         \begin{minipage} [b] {0.70\textwidth}
        \begin{minipage}[c]{0.57\textwidth}
        \centering
        \subfigure[Query Recall]{
            \includegraphics[width=0.99\textwidth]{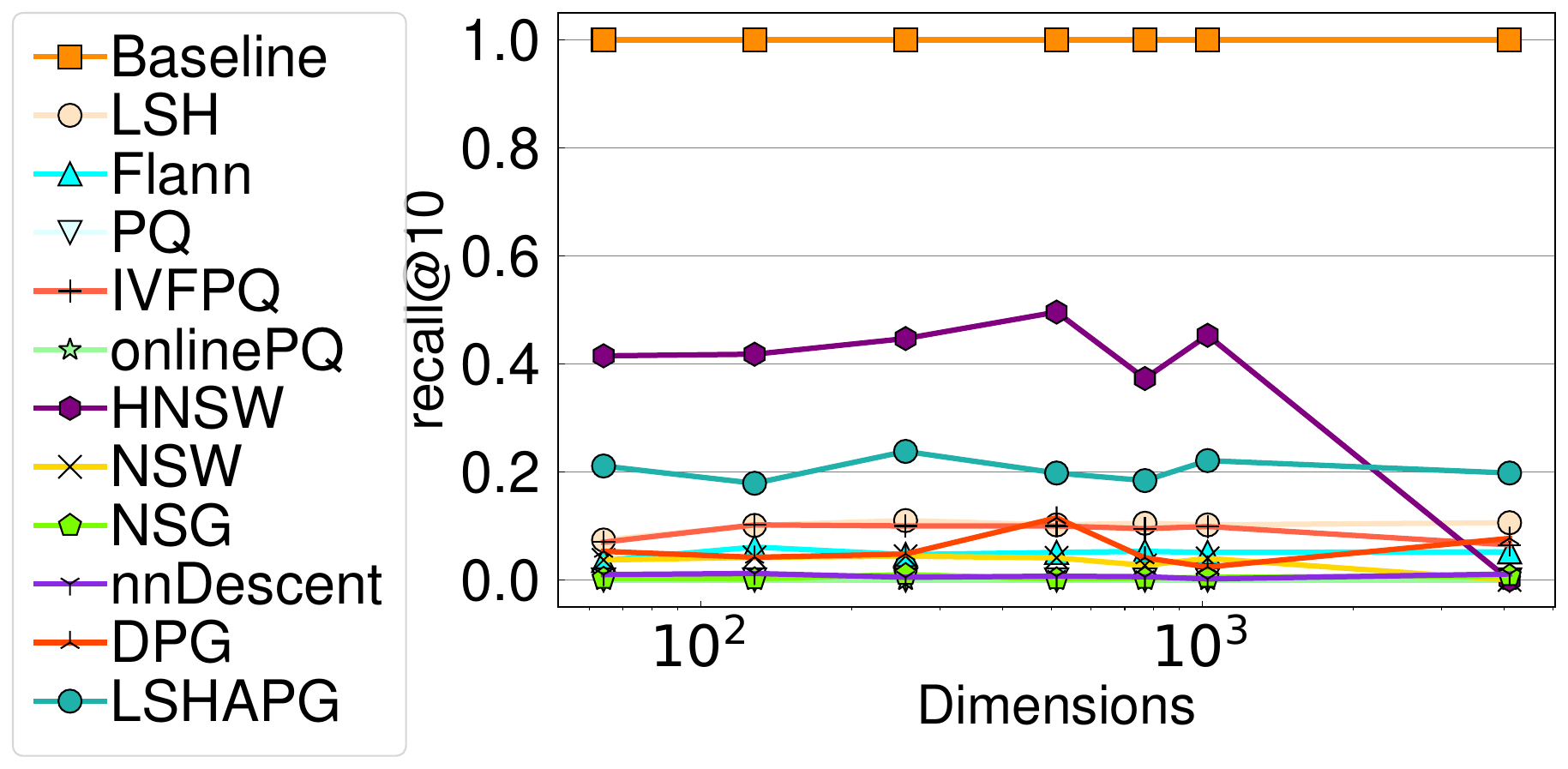}
            \label{fig:dim_recall}
        }
    \end{minipage}
    \begin{minipage}[c]{0.41\textwidth}
        \subfigure[Query Latency.]{
            \includegraphics[width=0.99\textwidth]{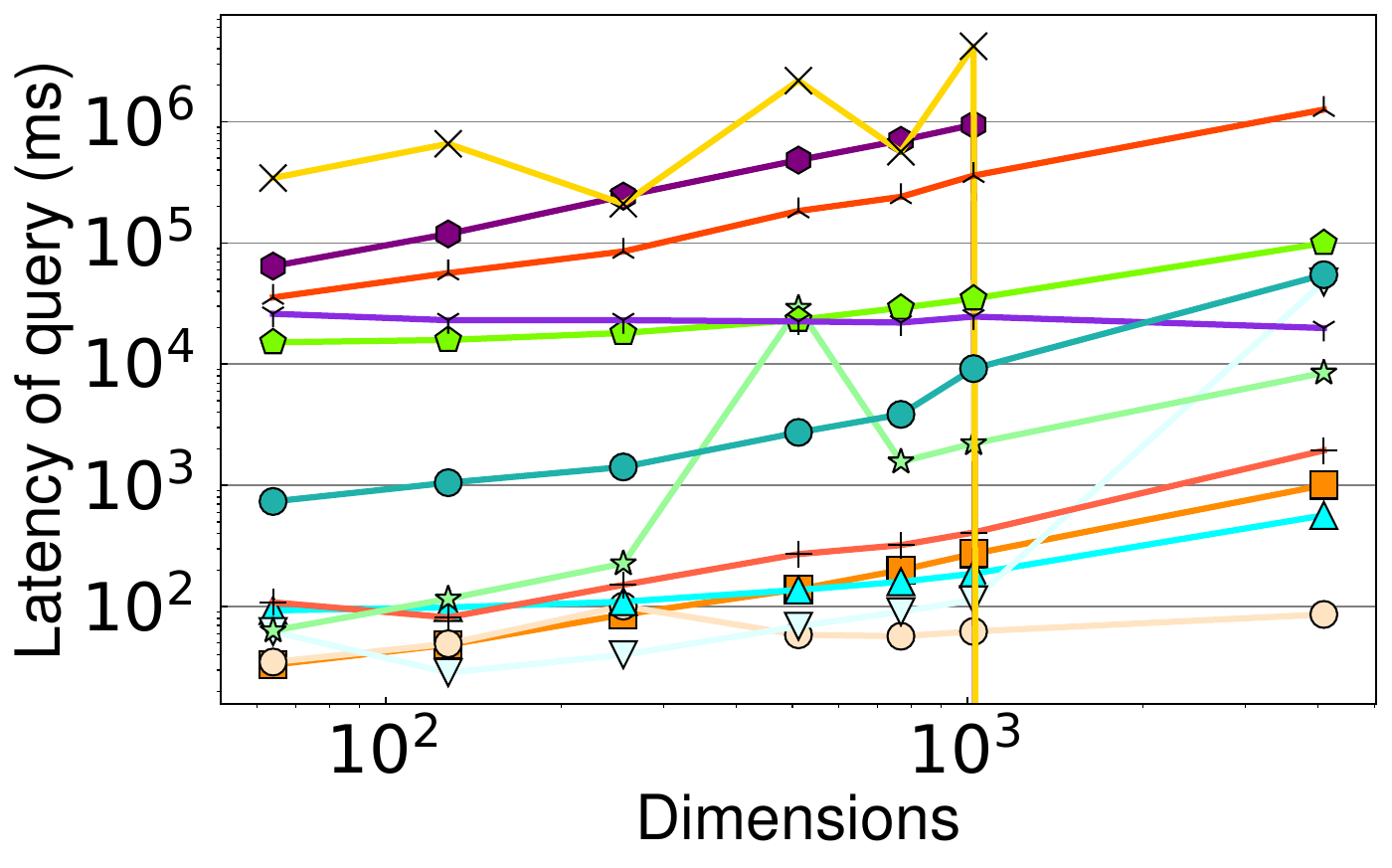}
            \label{fig:dim_lat}
        }
    \end{minipage}     
    \caption{Revisit the ``curse of dimension'' under online ingestion. \algoHNSW and \algoNSW run out of memory under 4096-D case.}
        \label{fig:dim}
    \end{minipage}
    
\end{minipage}
\end{figure}

\begin{figure}[t]
\begin{minipage}[a]{\textwidth} 
\centering
         \begin{minipage} [b] {0.70\textwidth}
        \begin{minipage}[c]{0.57\textwidth}
        \centering
        \subfigure[Query Recall]{
            \includegraphics[width=0.99\textwidth]{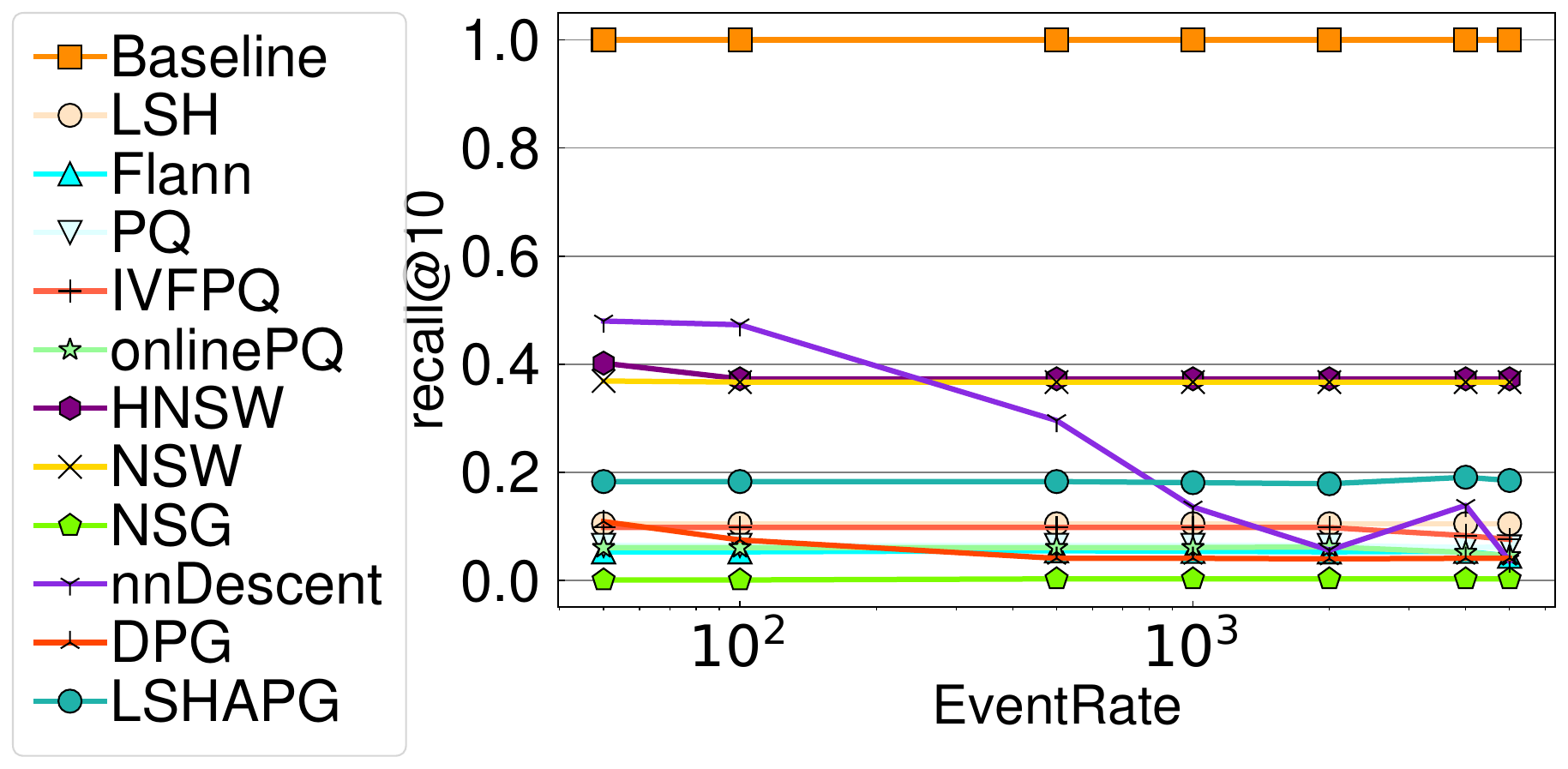}
            \label{fig:rate_recall}
        }
    \end{minipage}
    \begin{minipage}[c]{0.41\textwidth}
        \subfigure[Query Latency.]{
            \includegraphics[width=0.99\textwidth]{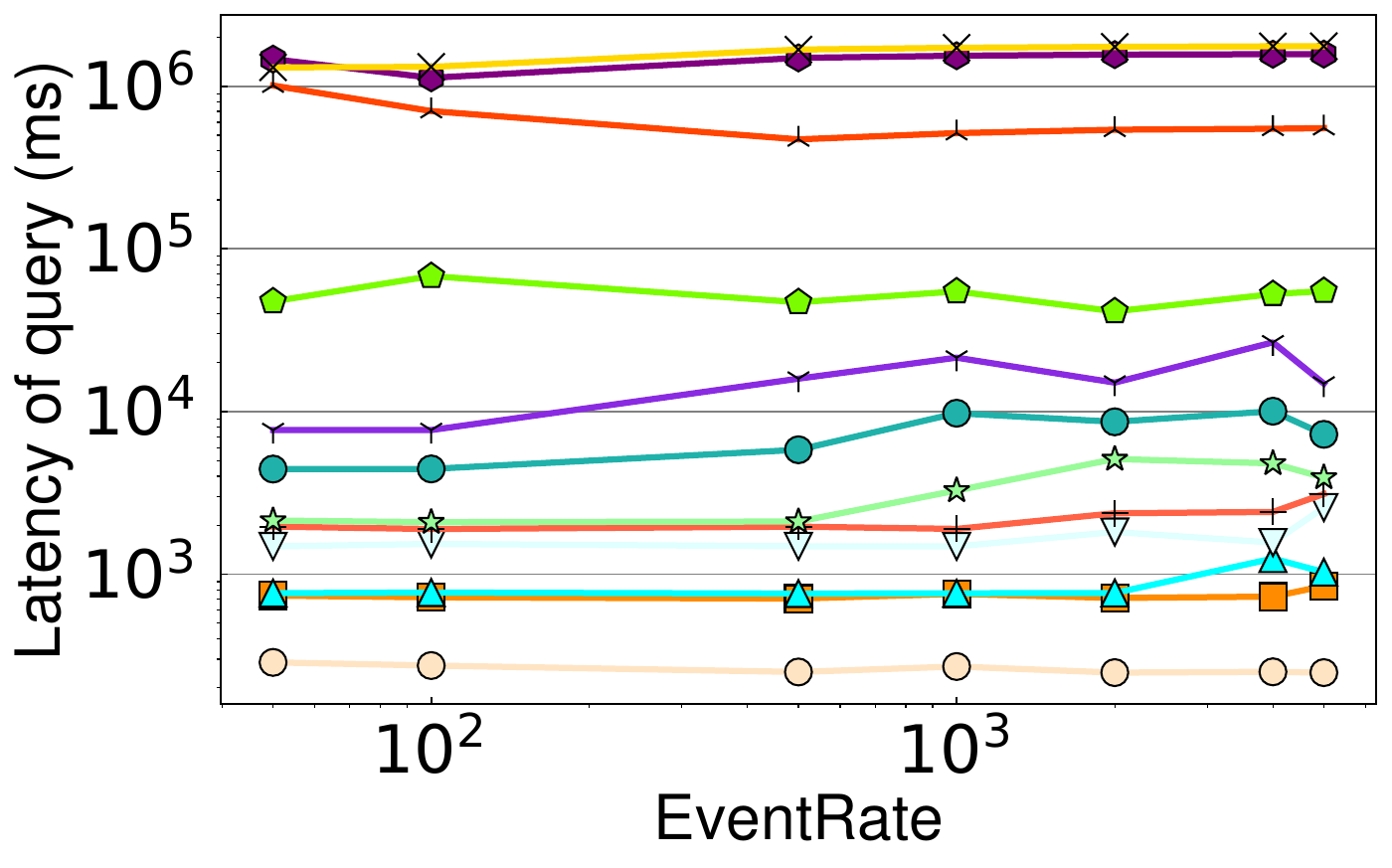}
            \label{fig:rate_lat}
        }
    \end{minipage}     
    \caption{Tuning event rate from $20$ to $5000$, the unit of event rate is $rows/s$.}
        \label{fig:event_rate}
    \end{minipage}
    
\end{minipage}
\end{figure}

\paragraph{Impact of Increased Event Rates on AKNN Performance} 
We examined the effects of varying event rates on AKNN performance using the \DPR dataset, adjusting the data rate in \system from 20 to 5000 rows per second. As shown in Figure~\ref{fig:event_rate}, AKNN systems generally perform worse as the event rate surpasses their ingestion capacity—for instance, about 20 rows per second for \algoHNSW—resulting in increased query latency and decreased recall due to data loss. Although \algoLSH manages to keep pace with an event rate of 5000 rows/s, its recall remains too low for practical use. We noted that navigation-based AKNNs like \algoHNSW, \algoNSW, and \algoNSG exhibit much greater fluctuations in query latency—up to ten times higher—compared to ranging-based AKNNs such as \algoLSH or \algoIVFPQ. This instability stems from the iterative scanning and comparison requirements within their proximity graphs, making them more susceptible to mismatches between data arrival and ingestion rates.

\begin{figure}[t]
\begin{minipage}[a]{\textwidth} 
\centering
         \begin{minipage} [b] {0.70\textwidth}
        \begin{minipage}[c]{0.57\textwidth}
        \centering
        \subfigure[Query Recall]{
            \includegraphics[width=0.99\textwidth]{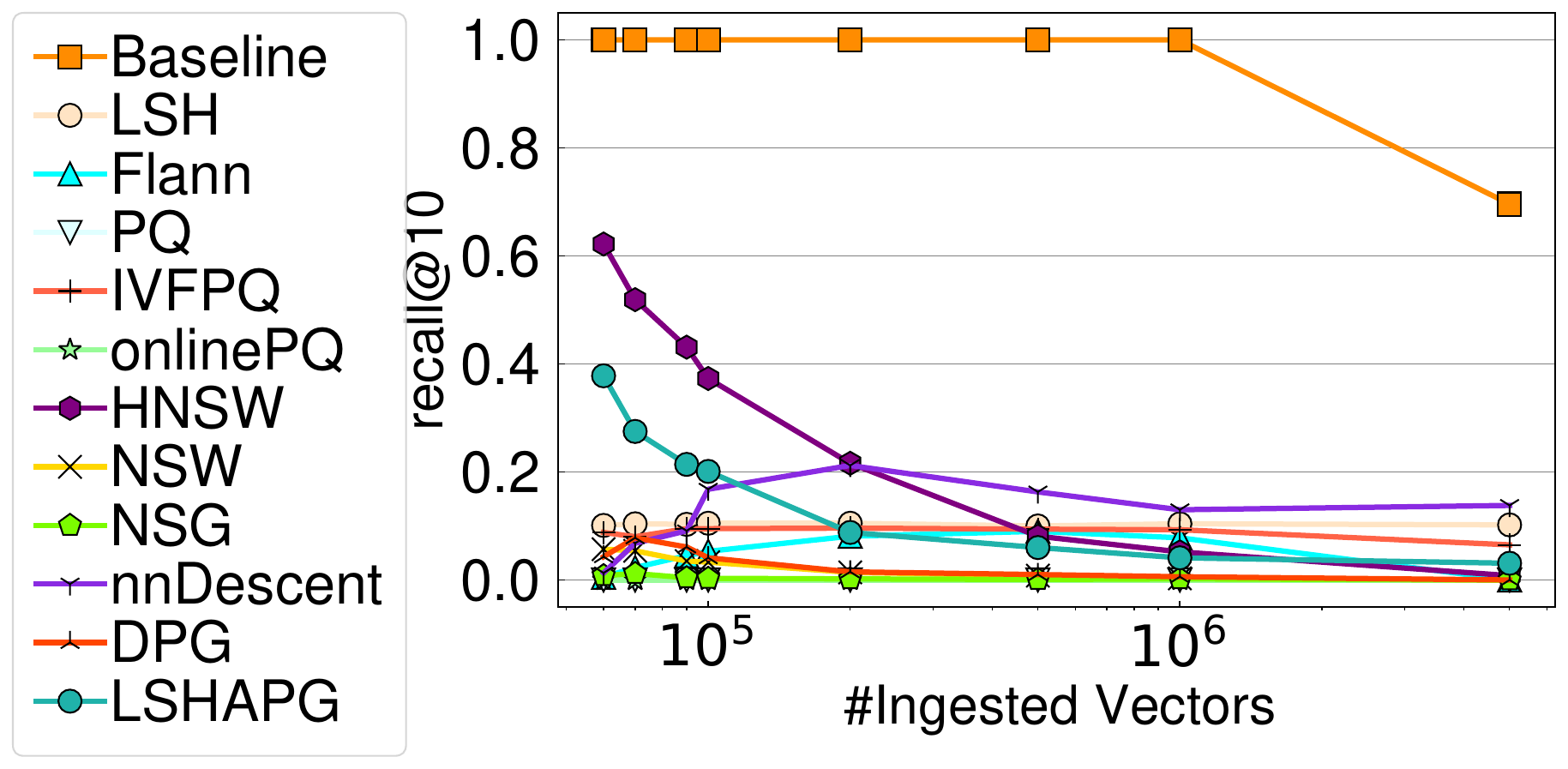}
            \label{fig:vecVol_recall}
        }
    \end{minipage}
    \begin{minipage}[c]{0.41\textwidth}
        \subfigure[Query Latency.]{
            \includegraphics[width=0.99\textwidth]{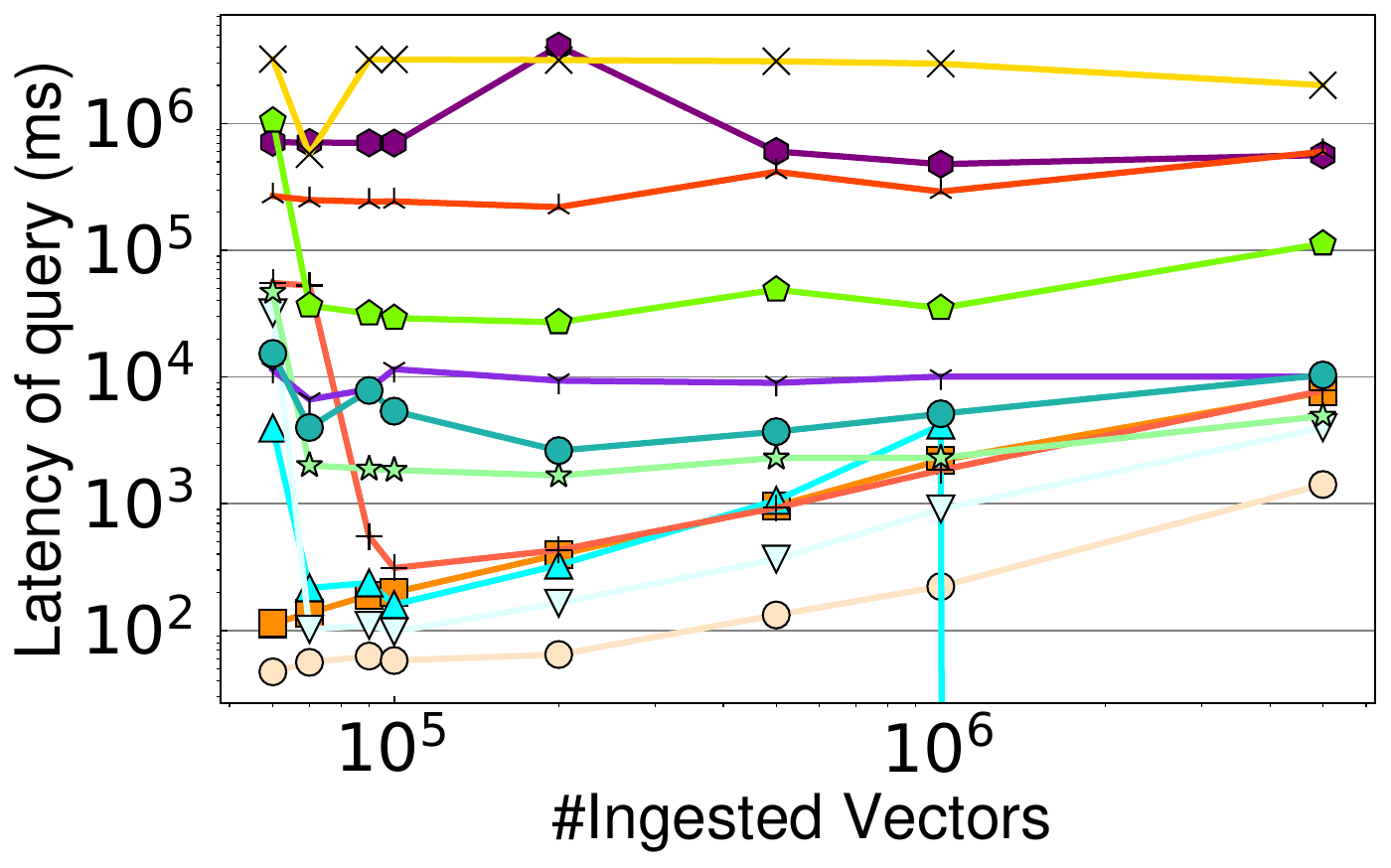}
            \label{fig:vecVol_lat}
        }
    \end{minipage}     
    \caption{Tuning the vector volume from $60K \sim 5M$. \algoFlann runs out of memory when the volume reaches $5M$.}
        \label{fig:vecVol}
    \end{minipage}
    
\end{minipage}
\end{figure}

\paragraph{Impact of Data Volume on AKNN Performance Under Online Ingestion}
Data scalability related to vector volume remains a crucial challenge for Approximate AKNN, as noted in prior research~\cite{li2019approximate}. To assess this in an online ingestion context, we used the \Random dataset with vector volumes ranging from 60K to 5M vectors, while the first 50K vectors were pre-loaded during the offline phase. Figure~\ref{fig:vecVol} illustrates how these adjustments affect query recall and latency. Our findings reveal two main points: First, consistent with prior studies~\cite{li2019approximate, xu2018online}, there is a general increase in query latency as vector volume grows, with \algoOnlinePQ showing a relatively slower rate of increase compared to \algoBF, suggesting some level of efficiency in managing larger data volumes (Figure~\ref{fig:vecVol_lat}).

Second, this dynamic ingestion scenario complicates the accuracy outcomes for AKNN. Although \algoHNSW typically shows superior accuracy, its performance drops below \algoLSH when handling more than 500K vectors (Figure~\ref{fig:vecVol_recall}). This decline contradicts previous static evaluations where \algoHNSW maintained high recall regardless of data volume~\cite{malkov2018efficient}. The observed decrease in accuracy is due to inefficient data ingestion and the resultant data dropping, which impedes the algorithm's ability to maintain accurate vector searches. Similarly, \algoBF begins to lose accuracy at a vector volume of 5M, highlighting the operational costs associated with reallocating and copying large data volumes during ingestion, even when managed in-memory.

%% file: Evaluations/Setups.tex
\subsection{Additional Setup Details for Machine Learning and Distance Computation Optimization}
\label{subsec:eva_setup}
For machine learning optimization, we employ a neural network (NN) to replace the traditional hash function in \algoLSH, enhancing its data organization capability. The NN, a simple multilayer perceptron (MLP), is pre-trained offline. It consists of a linear encoding layer with a \emph{tanh} activation function, followed by a decoding layer, tailored to approximate the spectral hashing loss function~\cite{NIPS2008_d58072be}. The output undergoes an adjustment by subtracting a trainable bias matrix for refined hashing. This implementation focuses on computational simplicity and generalization to typical AKNN tasks.

For distance compute optimization, we implement distance optimization strategies for \algoHNSW using L2 distance instead of the default inner product to align with specific optimization techniques~\cite{gao2023high}. The optimizations include: 1) Locally-Adaptive Vector Quantization (LVQ) \cite{aguerrebere2024locally}, which uses a global mean to scale vectors and local extremes for quantization, thus allowing for reduced computation costs in distance calculations; 2) ADSampling \cite{gao2023high}, which introduces a randomized transformation of input and query vectors, with incremental distance evaluations against a pre-defined threshold to potentially cut off unnecessary computations, thus enhancing processing speed and efficiency.

%% file: neurips_2023.bbl
\begin{thebibliography}{10}

\bibitem{tweetsSpeed}
The number of tweets per day in 2022 , \url{https://www.dsayce.com/category/social-media/}, 2022.
\newblock Last Accessed: 2023-12-05.

\bibitem{warthunderWiki}
War thunder wiki , \url{https://wiki.warthunder.com/}, 2024.
\newblock Last Accessed: 2024-04-05.

\bibitem{aguerrebere2024locally}
C.~Aguerrebere, M.~Hildebrand, I.~S. Bhati, T.~Willke, and M.~Tepper.
\newblock Locally-adaptive quantization for streaming vector search.
\newblock {\em arXiv preprint arXiv:2402.02044}, 2024.

\bibitem{andrews2004scheduling}
M.~Andrews, K.~Kumaran, K.~Ramanan, A.~Stolyar, R.~Vijayakumar, and P.~Whiting.
\newblock Scheduling in a queuing system with asynchronously varying service rates.
\newblock {\em Probability in the Engineering and Informational Sciences}, 18(2):191--217, 2004.

\bibitem{arun2018copa}
V.~Arun and H.~Balakrishnan.
\newblock Copa: Practical $\{$Delay-Based$\}$ congestion control for the internet.
\newblock In {\em 15th USENIX Symposium on Networked Systems Design and Implementation (NSDI 18)}, pages 329--342, 2018.

\bibitem{baranchuk2019learning}
D.~Baranchuk, D.~Persiyanov, A.~Sinitsin, and A.~Babenko.
\newblock Learning to route in similarity graphs.
\newblock In {\em International Conference on Machine Learning}, pages 475--484. PMLR, 2019.

\bibitem{cidon2015tiered}
A.~Cidon, R.~Escriva, S.~Katti, M.~Rosenblum, and E.~G. Sirer.
\newblock Tiered replication: A cost-effective alternative to full cluster geo-replication.
\newblock In {\em 2015 USENIX Annual Technical Conference (USENIX ATC 15)}, pages 31--43, 2015.

\bibitem{dong2003concept}
A.~Dong and B.~Bhanu.
\newblock Concept learning and transplantation for dynamic image databases.
\newblock In {\em 2003 International Conference on Multimedia and Expo. ICME'03. Proceedings (Cat. No. 03TH8698)}, volume~1, pages I--765. IEEE, 2003.

\bibitem{dong2011efficient}
W.~Dong, C.~Moses, and K.~Li.
\newblock Efficient k-nearest neighbor graph construction for generic similarity measures.
\newblock In {\em Proceedings of the 20th international conference on World wide web}, pages 577--586, 2011.

\bibitem{fowler1991local}
H.~J. Fowler and W.~E. Leland.
\newblock Local area network characteristics, with implications for broadband network congestion management.
\newblock {\em IEEE Journal on Selected Areas in Communications}, 9(7):1139--1149, 1991.

\bibitem{fu2019fast}
C.~Fu, C.~Xiang, C.~Wang, and D.~Cai.
\newblock Fast approximate nearest neighbor search with the navigating spreading-out graph.
\newblock {\em Proceedings of the VLDB Endowment}, 12(5):461--474, 2019.

\bibitem{gao2023high}
J.~Gao and C.~Long.
\newblock High-dimensional approximate nearest neighbor search: with reliable and efficient distance comparison operations.
\newblock {\em Proceedings of the ACM on Management of Data}, 1(2):1--27, 2023.

\bibitem{gionis1999similarity}
A.~Gionis, P.~Indyk, and R.~Motwani.
\newblock Similarity search in high dimensions via hashing.
\newblock In {\em VLDB}, volume~99, pages 518--529, 1999.

\bibitem{heliou2020gradient}
A.~H{\'e}liou, P.~Mertikopoulos, and Z.~Zhou.
\newblock Gradient-free online learning in continuous games with delayed rewards.
\newblock In {\em International conference on machine learning}, pages 4172--4181. PMLR, 2020.

\bibitem{huang2019gpipe}
Y.~Huang, Y.~Cheng, A.~Bapna, O.~Firat, D.~Chen, M.~Chen, H.~Lee, J.~Ngiam, Q.~V. Le, Y.~Wu, et~al.
\newblock Gpipe: Efficient training of giant neural networks using pipeline parallelism.
\newblock {\em Advances in neural information processing systems}, 32, 2019.

\bibitem{indyk1998approximate}
P.~Indyk and R.~Motwani.
\newblock Approximate nearest neighbors: towards removing the curse of dimensionality.
\newblock In {\em Proceedings of the thirtieth annual ACM symposium on Theory of computing}, pages 604--613. ACM, 1998.

\bibitem{jegou2010product}
H.~Jegou, M.~Douze, and C.~Schmid.
\newblock Product quantization for nearest neighbor search.
\newblock {\em IEEE transactions on pattern analysis and machine intelligence}, 33(1):117--128, 2010.

\bibitem{jiang2023chameleon}
W.~Jiang, M.~Zeller, R.~Waleffe, T.~Hoefler, and G.~Alonso.
\newblock Chameleon: a heterogeneous and disaggregated accelerator system for retrieval-augmented language models.
\newblock {\em arXiv preprint arXiv:2310.09949}, 2023.

\bibitem{karpukhin2020dense}
V.~Karpukhin, B.~O{\u{g}}uz, S.~Min, P.~Lewis, L.~Wu, S.~Edunov, D.~Chen, and W.-t. Yih.
\newblock Dense passage retrieval for open-domain question answering.
\newblock {\em arXiv preprint arXiv:2004.04906}, 2020.

\bibitem{lewis2020retrieval}
P.~Lewis, E.~Perez, A.~Piktus, F.~Petroni, V.~Karpukhin, N.~Goyal, H.~K{\"u}ttler, M.~Lewis, W.-t. Yih, T.~Rockt{\"a}schel, et~al.
\newblock Retrieval-augmented generation for knowledge-intensive nlp tasks.
\newblock {\em Advances in Neural Information Processing Systems}, 33:9459--9474, 2020.

\bibitem{li2022deep}
M.~Li, Y.-G. Wang, P.~Zhang, H.~Wang, L.~Fan, E.~Li, and W.~Wang.
\newblock Deep learning for approximate nearest neighbour search: A survey and future directions.
\newblock {\em IEEE Transactions on Knowledge and Data Engineering}, 2022.

\bibitem{li2023learning}
W.~Li, C.~Feng, D.~Lian, Y.~Xie, H.~Liu, Y.~Ge, and E.~Chen.
\newblock Learning balanced tree indexes for large-scale vector retrieval.
\newblock In {\em Proceedings of the 29th ACM SIGKDD Conference on Knowledge Discovery and Data Mining}, pages 1353--1362, 2023.

\bibitem{li2019approximate}
W.~Li, Y.~Zhang, Y.~Sun, W.~Wang, M.~Li, W.~Zhang, and X.~Lin.
\newblock Approximate nearest neighbor search on high dimensional data—experiments, analyses, and improvement.
\newblock {\em IEEE Transactions on Knowledge and Data Engineering}, 32(8):1475--1488, 2019.

\bibitem{loper2002nltk}
E.~Loper and S.~Bird.
\newblock Nltk: The natural language toolkit.
\newblock {\em arXiv preprint cs/0205028}, 2002.

\bibitem{luo2022semantic}
X.~Luo, H.-H. Chen, and Q.~Guo.
\newblock Semantic communications: Overview, open issues, and future research directions.
\newblock {\em IEEE Wireless Communications}, 29(1):210--219, 2022.

\bibitem{lyu2024crud}
Y.~Lyu, Z.~Li, S.~Niu, F.~Xiong, B.~Tang, W.~Wang, H.~Wu, H.~Liu, T.~Xu, and E.~Chen.
\newblock Crud-rag: A comprehensive chinese benchmark for retrieval-augmented generation of large language models.
\newblock {\em arXiv preprint arXiv:2401.17043}, 2024.

\bibitem{malkov2014approximate}
Y.~Malkov, A.~Ponomarenko, A.~Logvinov, and V.~Krylov.
\newblock Approximate nearest neighbor algorithm based on navigable small world graphs.
\newblock {\em Information Systems}, 45:61--68, 2014.

\bibitem{malkov2018efficient}
Y.~A. Malkov and D.~A. Yashunin.
\newblock Efficient and robust approximate nearest neighbor search using hierarchical navigable small world graphs.
\newblock In {\em IEEE Transactions on Pattern Analysis and Machine Intelligence}. IEEE, 2018.

\bibitem{manning2008introduction}
C.~D. Manning, P.~Raghavan, and H.~Sch{\"u}tze.
\newblock Introduction to information retrieval.
\newblock 2008.

\bibitem{moritz2018ray}
P.~Moritz, R.~Nishihara, S.~Wang, A.~Tumanov, R.~Liaw, E.~Liang, M.~Elibol, Z.~Yang, W.~Paul, M.~I. Jordan, et~al.
\newblock Ray: A distributed framework for emerging $\{$AI$\}$ applications.
\newblock In {\em 13th USENIX symposium on operating systems design and implementation (OSDI 18)}, pages 561--577, 2018.

\bibitem{muja2014scalable}
M.~Muja and D.~G. Lowe.
\newblock Scalable nearest neighbor algorithms for high dimensional data.
\newblock {\em IEEE transactions on pattern analysis and machine intelligence}, 36(11):2227--2240, 2014.

\bibitem{Pytorch}
NA.
\newblock Pytorch homepage, \url{https://pytorch.org/}, 2023.

\bibitem{singh2021freshdiskann}
A.~Singh, S.~J. Subramanya, R.~Krishnaswamy, and H.~V. Simhadri.
\newblock Freshdiskann: A fast and accurate graph-based ann index for streaming similarity search.
\newblock {\em arXiv preprint arXiv:2105.09613}, 2021.

\bibitem{sundaram2013streaming}
N.~Sundaram, A.~Turmukhametova, N.~Satish, T.~Mostak, P.~Indyk, S.~Madden, and P.~Dubey.
\newblock Streaming similarity search over one billion tweets using parallel locality-sensitive hashing.
\newblock {\em Proceedings of the VLDB Endowment}, 6(14):1930--1941, 2013.

\bibitem{wang2021milvus}
J.~Wang, X.~Yi, R.~Guo, H.~Jin, P.~Xu, S.~Li, X.~Wang, X.~Guo, C.~Li, X.~Xu, et~al.
\newblock Milvus: A purpose-built vector data management system.
\newblock In {\em Proceedings of the 2021 International Conference on Management of Data}, pages 2614--2627, 2021.

\bibitem{wang2021comprehensive}
M.~Wang, X.~Xu, Q.~Yue, and Y.~Wang.
\newblock A comprehensive survey and experimental comparison of graph-based approximate nearest neighbor search.
\newblock {\em Proceedings of the VLDB Endowment}, 14(11):1964--1978, 2021.

\bibitem{weber1998quantitative}
R.~Weber, H.-J. Schek, and S.~Blott.
\newblock A quantitative analysis and performance study for similarity-search methods in high-dimensional spaces.
\newblock {\em VLDB}, 98:194--205, 1998.

\bibitem{NIPS2008_d58072be}
Y.~Weiss, A.~Torralba, and R.~Fergus.
\newblock Spectral hashing.
\newblock In D.~Koller, D.~Schuurmans, Y.~Bengio, and L.~Bottou, editors, {\em Advances in Neural Information Processing Systems}, volume~21. Curran Associates, Inc., 2008.

\bibitem{xu2018online}
D.~Xu, I.~W. Tsang, and Y.~Zhang.
\newblock Online product quantization.
\newblock {\em IEEE Transactions on Knowledge and Data Engineering}, 30(11):2185--2198, 2018.

\bibitem{yuan2022decentralized}
B.~Yuan, Y.~He, J.~Davis, T.~Zhang, T.~Dao, B.~Chen, P.~S. Liang, C.~Re, and C.~Zhang.
\newblock Decentralized training of foundation models in heterogeneous environments.
\newblock {\em Advances in Neural Information Processing Systems}, 35:25464--25477, 2022.

\bibitem{zeng2024pecj}
X.~Zeng, S.~Zhang, H.~Zhong, H.~Zhang, M.~Lu, Z.~Zheng, and Y.~Chen.
\newblock Pecj: Stream window join on disorder data streams with proactive error compensation.
\newblock {\em Proceedings of the ACM on Management of Data}, 2(1):1--24, 2024.

\bibitem{zhang2023scalable}
H.~Zhang, X.~Zeng, S.~Zhang, X.~Liu, M.~Lu, and Z.~Zheng.
\newblock Scalable online interval join on modern multicore processors in openmldb.
\newblock In {\em 2023 IEEE 39th International Conference on Data Engineering (ICDE)}, pages 3031--3042. IEEE, 2023.

\bibitem{zhang2021parallelizing}
S.~Zhang, Y.~Mao, J.~He, P.~M. Grulich, S.~Zeuch, B.~He, R.~T. Ma, and V.~Markl.
\newblock Parallelizing intra-window join on multicores: An experimental study.
\newblock In {\em Proceedings of the 2021 International Conference on Management of Data}, pages 2089--2101, 2021.

\bibitem{zhao2023towards}
X.~Zhao, Y.~Tian, K.~Huang, B.~Zheng, and X.~Zhou.
\newblock Towards efficient index construction and approximate nearest neighbor search in high-dimensional spaces.
\newblock {\em Proceedings of the VLDB Endowment}, 16(8):1979--1991, 2023.

\end{thebibliography}
